\documentclass{article}
\usepackage{arxiv}
\usepackage[utf8]{inputenc}
\usepackage[T1]{fontenc}    
\usepackage{graphicx}
\usepackage{subfig}       
\graphicspath{{figures/}} 
\usepackage{psfrag}       
\usepackage{amsmath,bm,amsfonts,mathrsfs}   
\usepackage{shortcuts}    
\usepackage{color}
\usepackage{hyperref}       
\usepackage{url}            
\usepackage{enumitem}
\usepackage{amsmath}
\usepackage{algorithm,algorithmic}
\usepackage{stmaryrd}
\usepackage{longtable}

\usepackage[color=green!20]{todonotes}

\usepackage{comment}

\usepackage{scalerel}
\DeclareMathOperator*{\foo}{\scalerel*{A}{\sum}}
\DeclareMathOperator*{\barr}{\scalerel*{A}{\textstyle\sum}}

\def\Balpha{\mbox{\boldmath$\alpha$}}
\def\Bbeta{\mbox{\boldmath$\beta$}}

\def\Bxi{\mbox{\boldmath$\xi$}}

\def\Btau{\mbox{\boldmath$\tau$}}

\def\Bvarepsilon{\mbox{\boldmath$\varepsilon$}}

\def\Bvarphi{\mbox{\boldmath$\varphi$}}

\title{A treatment of particle-electrolyte sharp interface fracture in solid-state batteries with multi-field discontinuities}

\author{
  ~Xiaoxuan Zhang$^1$, ~Tryaksh Gupta$^{1,3}$, ~Zhenlin Wang$^1$, ~Amalie Trewartha$^4$, ~Abraham Anapolsky$^4$, ~Krishna Garikipati$^{1,2,3}$ \thanks{Corresponding author. E-mail address: krishna@umich.edu \hfill \today} \\[3mm]
   $^1$Department of Mechanical Engineering, University of Michigan, United States \\
   $^2$Department of Mathematics, University of Michigan, United States \\
   $^3$Michigan Institute for Computational Discovery \& Engineering, University of Michigan, United States \\
   $^4$ Toyota Research Institute, United States
}

\begin{document}

\maketitle

\begin{abstract}
  In this work, we present a computational  framework for coupled electro-chemo-(nonlinear) mechanics at the particle scale for solid-state batteries. The framework accounts for   interfacial fracture between the  active particles and solid electrolyte due to  intercalation stresses. We extend discontinuous finite element methods for a sharp interface treatment of discontinuities in concentrations, fluxes, electric fields and in displacements, the latter arising from active particle-solid electrolyte interface fracture. We model the  degradation in the charge transfer process that results from the loss of contact due to fracture at  the electrolyte-active particle interfaces.
  Additionally, we account for the stress-dependent kinetics that can influence the charge transfer reactions and solid state diffusion. 
  The discontinuous finite element approach does not require a conformal mesh. 
  This offers the flexibility to construct arbitrary particle shapes and geometries that are based on design, or are obtained from microscopy images. The finite element mesh, however, can remain Cartesian, and independent of the particle geometries.  
  We demonstrate this computational framework on micro-structures that are representative of solid-sate batteries with single and multiple anode and cathode particles. 
\end{abstract}


\section{Introduction}

Solid-state batteries (SSBs) are gaining in interest due to their high energy density and  improved safety over liquid electrolyte-based systems. However, the development of SSBs also faces challenges. Some of them such as dendrite formation or interphases growing at solid electrolyte-active particle interfaces are the driven by complex electrochemical coupling \cite{lewis2019chemo}. However, several others stem from higher mechanical stresses relative to liquid electrolyte systems, which develop in the all solid system of electrode, electrolyte, binder and current collector.  In most SSB chemistries, the stress arises from intercalation strains and the confined all-solid battery configuration. As the strains cycle with charge and discharge, so do the stresses, and can lead to many of the failure phenomena that are well-known in solid mechanics.

Possibly the most important of these is fracture, which can arise at many locations including the active cathode and anode particles, and brittle electrolytes such as the ceramics $\beta\text{-Li}_3\text{PS}_4$ or lithium lanthanum zirconium oxides (LLZO). By its being an interface, the electrode particle-electrolyte junction is particularly susceptible to fracture. New surfaces are created at this interface with a loss of contact. Reactions, which rely upon proximity between components, can be compromised by the introduction of physical gaps between materials on either side of the fracture surfaces. This can become a major concern for the solid electrolyte-active particle interface, which, of course, is critical for charge transfer. The suppression of charge transfer across this fractured interface causes a degradation in capacity, which over many charge-discharge cycles can lead to loss of stability of electrochemical performance. The details of this coupling between mechanics and electrochemistry, specifically how the charge transfer kinetics falls off with growth in the fracture-induced gap remains poorly understood. However, it could be modelled phenomenologically, and allow access to computational studies of the loss in electrochemical performance with interface crack growth over charge-discharge cycles. 

This is a multiphysics problem: (a) Intercalation strains drive the stress. Depending upon the chemistry, there is local shrinkage of the cathode or anode particle during lithiation or delithiation (assuming a lithium battery). (b) The tensile stress during particle shrinkage can cause  fracture of the solid electrolyte-cathode/anode interface. (c) The change in charge transport due to loss in contact at the interface either transfers less lithium to the cathode/anode or extracts less during charge/discharge, all depending on the cathode/anode chemistry. (d) The loss of load-carrying capacity across the fractured interface leads to a lowering of the tensile stress throughout the solid state battery. The altered charge transfer, and therefore (de)lithiation-driven (de)intercalation also changes the stress as cycles progress. (e) There is an additional effect of stress-mediated kinetic phenomena: on Li and Li$^+$ transport in the active particles and electrolyte, respectively, and on charge transfer reactions across the interface. As the stress changes with cycles so do these stress-mediated kinetics.

Over hundreds of cycles, there is a progressive capacity loss, that is often manifested by a steep degradation past some threshold. The complexities of this process make it difficult to gain insight purely from experiment. Multiphysics modelling and computation are indispensable, and have led to a number of lines of investigation. In a pair of papers, Klinsmann et al. developed a phase field fracture-based model for crack growth during extraction \cite{klinsmann2015modeling} and insertion \cite{klinsmann2016modeling} of Li in active particles. Their models included diffusive transport, linearized mechanics and the phase field formulation of damage and fracture in a coupled treatment, but without the electric fields. The electrolytes were not solid state, and therefore intra-particle fracture was of interest to the authors rather than particle-electrolyte interface fracture. Nevertheless, these works mark an important step toward the treatment of intercalation stress-induced fracture. Ganser et al. presented a free energy-based coupled treatment of electrostatics, mass/charge transport and nonlinear elasticity for fully solid state batteries \cite{ganser2019finite}. The electrolyte in their numerical models was a solid polymer. The same group of authors used this model to study how the elastic properties of the solid polymer electrolyte influence the stability of its interface with metal active particles to perturbations that could grow into protrusions \cite{ganser2020stiffer,ganser2021electro}. Debonding at the active particle-binder interface  was treated by Iqbal et al. using cohesive zone elements, a chemo-mechanical model with linearized elasticity for the particle and Neo-Hookean elasticity for the binder \cite{iqbal2021progressive}. Rezaei et al. also carried out a similar treatment of chemo-mechanically driven  fracture in solid state batteries using cohesive zone models \cite{rezaei2021consistent}. Also related is these authors' extension of this model to phase field fracture to study active particle fracture \cite{rezaei2023cohesive}. Electrolyte, intra-particle and active particle-solid electrolyte interface fracture were modelled using linearized elasticity, damage and cohesive zone elements, and the accumulation of damage (degradation) with cycling was demonstrated. Bistri and Di Leo developed a novel surface element to resolve the chemo-mechanics at the active particle-electrolyte interface \cite{bistri2021amodeling}. Of interest in their work is the modelling of multi-particle configurations and their relation to the development of capacity loss. 

Other work in the literature has accounted for the influence of mechanics on the charge transfer kinetics at the active particle-electrolyte interface. Ganser et al. worked within the framework of transition state theory to propose extensions of the classical Butler-Volmer model \cite{ganser2019extended}. Zhao et al. provided an electro-chemo-mechanical treatment considering void formation and growth at active particle-solid electrolyte interfaces. The elasto-viscoplastic response of Li was accounted for, with phase field models for the formation of voids from vacancies \cite{zhao2022phase}. Afshar and Di Leo have presented a thermodynamically based chemo-mechanics treatment for resolving interface phenomena using phase field methods \cite{afshar2021thermodynamically}. A useful experimental study was carried out by Han et al., who performed cycling of Li NMC-anode, argyrodite-electrolyte solid state batteries, finding stresses in the mega Pascal range \cite{han2021stress}. This is an important marker for the ceramic electrolytes that are modelled in this communication.

Surveys of the the range of coupled phenomena and failure mechanisms in all solid state batteries have also appeared recently. These include the reviews by Bistri et al. \cite{bistri2021modeling}  and by Tian et al. \cite{tian2022review}, which focused on interface stability, interphase fracture and the chemo-mechanics of composite electrodes. A comprehensive review of the open questions in the coupled electro-chemo-mechanics of solid state batteries by Deshpande and McMeeking \cite{deshpande2023models} focused attention on the inadequacy of models that neglect the viscoplasticity of lithium metal electrodes, arguing for the importance of this effect on void formation and the growth of lithium  into cracks.
 
In this communication we propose an electro-chemo-mechanically coupled model to simulate SSBs by resolving the phenomena listed in (a-e) above at the scale of individual particles. Cathode/anode particle sizes and geometries naturally have strong influences on these physics, and it becomes important to carry out computational studies accounting for multi-particle configurations. Stiff solid electrolytes, such as the ceramics LLZO and even $\beta$-Li$_3$PS$_4$, in combination with stiff cathode particles such as Li NMC and graphite anodes lead to higher stresses. Fracture occurs with high probability at the active particle-solid electrolyte interface in such systems. With this motivation, a focus of this communication is on modeling fracture at the active particle-solid electrolyte interfaces. The modelling of interface fracture in realistic and experimental image-based microstructures with a distribution of particle shapes and sizes would be limited by meshes that conform to complicated microstructures by following electrolyte-particle interfaces. The creation of such meshes is a tedious and expensive undertaking, and could become a bottleneck as is well understood in computational engineering. To surmount this difficulty, we propose to use a scalar field to define each particle's location and geometry. The particle-electrolyte interface is implicitly modeled via the so-called the embedded interface method. Such a treatment allows the imposition of interface conditions on electrostatic, species density and deformation fields within two- or three-dimensional elements by extending discontinuous finite element methods \cite{armero1996analysis,rao2000modelling,garikipati2000variational,Garikipati2001Rao,garikipati2002variational,oliver2004continuum}. Thus, while our treatment bears clear similarities to the cohesive zone-based treatments of fracture discussed above \cite{iqbal2021progressive,rezaei2021consistent} and to the treatment that introduced surface elements for chemo-mechanics \cite{bistri2021amodeling}, the non-conforming meshes enabled by discontinuous finite elements allows greater flexibility for arbitrary multi-particle configurations. We exploit the natural treatment of discontinuous fields afforded by this approach to exactly account for the discontinuity of Li/Li$^+$ concentration fields and in the displacement field post-fracture at the particle-electrolyte interface. We additionally define distinct particle and electrolyte electric potentials and allow them to change discontinuously at the interface. We use this direct representation of discontinuities to drive interface charge transfer reactions and traction-displacement relations.

This work is organized as follows: 
In Section \ref{sec:theory}, we describe the standard governing equations, kinematics and constitutive relations of  the coupled electro-chemo-mechanics in solid-state batteries in the traditional setting--i.e., without accounting for interfaces. In doing so, we make connections to the derivations in Ganser et al. \cite{ganser2019finite}. In Section \ref{sec:interface} we introduce the treatment of interfaces in continuum physics, including the idea of implicitly  representing the particle's boundary with a scalar field. We also describe the numerical treatment of interface conditions, drawing upon finite element design.  In Section \ref{sec:meshgen}, we briefly describe our workflow for efficiently generating microstructures on a regular, Cartesian mesh based on the proposed approach. In Section \ref{sec:simulation}, we present the results of simulations  under a variety of coupling conditions for two- and multi-particle microstructures. In Section \ref{sec:conclusion}, we offer a discussion of our treatment, place it in context and suggest directions for its extension. 

\section{The electro-chemo-mechanics of solid state batteries}\label{sec:theory}
Newman's work has laid the foundation for modeling electrochemical systems and has been widely used in battery problems \cite{NewmanBook,newman1975porous,Newman1980,Newman1993,doyle1993modeling,fuller1994simulation}. In the past, we, among others, have extended this body of work to couple nonlinear mechanics with electrochemistry at both, the homogenized and particle-resolved scales  \cite{Wang+Garikipati2017, Wang+Garikipati2018}; however, our previous work has been for a system with a liquid electrolyte. We lay out the electro-chemo-mechanical problem for the case of a solid electrolyte. Rather than repeating the derivation from first principles that has been presented in the literature, we make connections to that work \cite{ganser2019finite}. 

\begin{figure}[h!]
    \centering
    \includegraphics[width=1.0\linewidth]{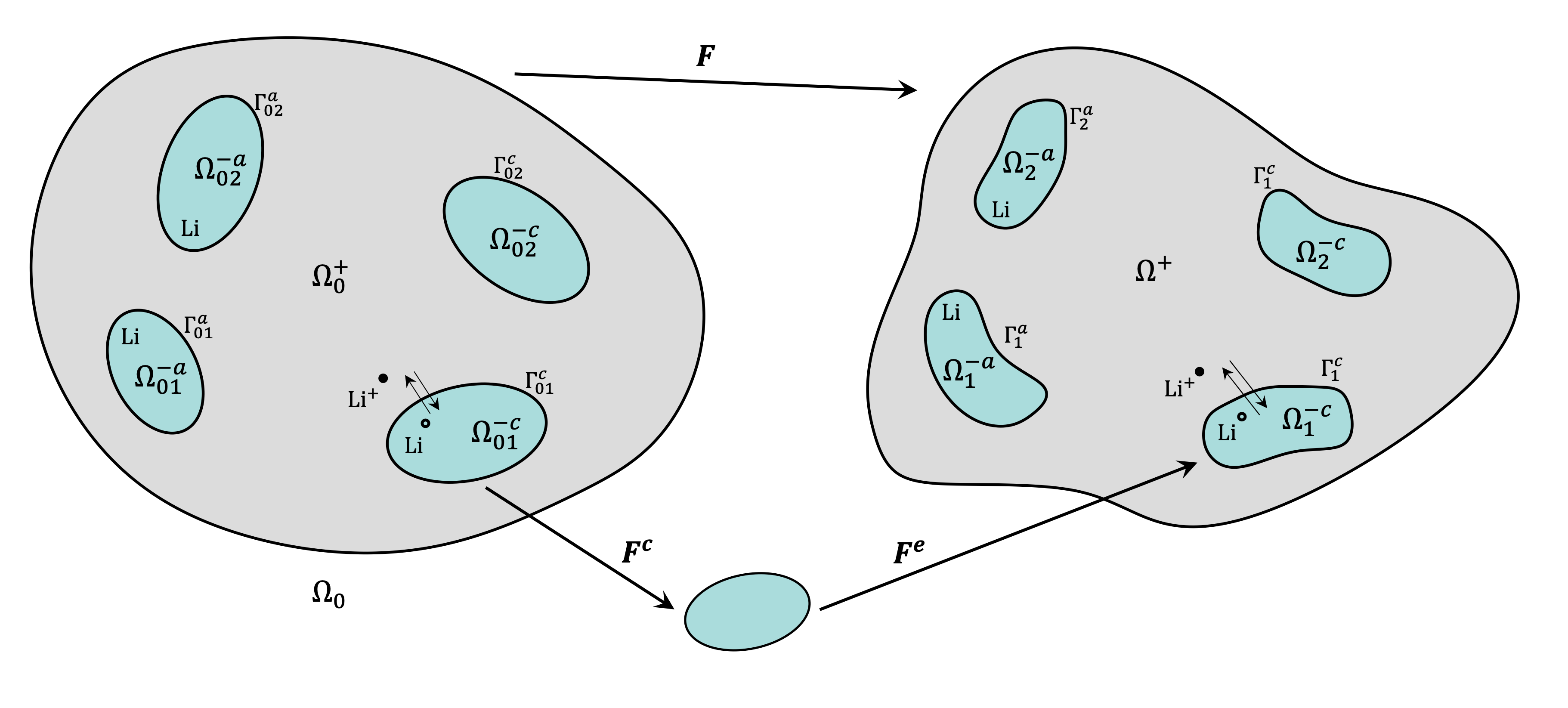}
    \caption{An illustration of the solid state battery in its reference configuration, $\Omega_0$ with solid electrolyte subdomain $\Omega_0^+$, anode particles $\Omega_{01}^{-\text{a}}, \Omega_{02}^{-\text{a}}$, cathode particles $\Omega_{01}^{-\text{c}}, \Omega_{02}^{-\text{c}}$ and interfaces $\Gamma_{01}^{\text{a}},\Gamma_{02}^{\text{a}},\Gamma_{01}^{\text{c}},\Gamma_{02}^{\text{c}}$. On the right is the deformed configuration. Also illustrated is the multiplicative decomposition: $\boldsymbol{F} = \boldsymbol{F}^\text{e}\boldsymbol{F}^\text{c}$.
    }
    \label{fig:deformation-map}
\end{figure}

We denote the continuum domain in its reference configuration by $\Omega_0$ (Figure \ref{fig:deformation-map}) and allow it to contain closed interfaces $\Gamma^\text{c}_{01},\dots \Gamma^\text{c}_{0m}$ and $\Gamma^\text{a}_{01},\dots \Gamma^\text{a}_{0n}$. Each $\Gamma^\text{c}_{0i},\; i  \in \{1,\dots,m\}$ is the boundary of an open subdomain $\Omega^{-\text{c}}_{0i}$ that represents a cathode particle; i.e., $\overline{\Omega^{-\text{c}}_{0i}} = \Omega^{-\text{c}}_{0i} \cup \Gamma^\text{c}_{0i}$and each $\Gamma^\text{a}_{0i},\; i  \in \{1,\dots,n\}$ is the boundary of an open subdomain $\Omega^{-\text{a}}_{0i}$ that represents an anode particle; i.e., $\overline{\Omega^{-\text{a}}_{0i}} = \Omega^{-\text{a}}_{0i} \cup \Gamma^\text{a}_{0i}$. The complement $\Omega_0\backslash\cup_{i=1}^m\overline{\Omega^{-\text{c}}_{0i}}\cup_{i=1}^n\overline{\Omega^{-\text{a}}_{0i}} = \Omega^+_0$. The simplest rendering of the solid state battery is with $\Omega^+_0$ being the multiply connected solid electrolyte subdomain, $\Omega^{-\text{c}}_{0i},\Omega^{-\text{a}}_{0j}$ being the cathode/anode particles and $\Gamma^\text{c}_{0i},\Gamma^\text{a}_{0j}$ being the corresponding cathode/anode-electrolyte interfaces. Additional sudomains of binders and current collectors will be made for numerical examples, but we avoid the tedious details here in the interest of brevity. Additionally, since the mathematical development is partly agnostic to the distinction between cathode and anode, we will use $\Omega^-_0$ for an active particle and $\Gamma_0$ as its interface, wherever the difference is inconsequential.

\subsection{Coupling conditions and governing equations}

\subsubsection{Lithiation, intercalation and kinematics}
The solid electrolyte $\Omega^+$ hosts Li$^+$ cations and the active particles, $\Omega^-$ are intercalated by Li. The discharge/charge reactions at the interfaces $\Gamma_0$ are:
\begin{subequations}
    \begin{align}
    \text{Discharge:}&\quad \text{Li} \rightarrow \text{Li}^+ + \text{e}^-\;\text{at } \Gamma^\text{a},\qquad \text{Li}^+ + \text{e}^- \rightarrow \text{Li}\;\text{at } \Gamma^\text{c}\label{eq:discharge}\\
    \text{Charge:}&\quad \text{Li}^+ + \text{e}^- \rightarrow \text{Li} \;\text{at } \Gamma^\text{a},\qquad \text{Li} \rightarrow \text{Li}^+ + \text{e}^- \;\text{at } \Gamma^\text{c}\label{eq:charge}
\end{align}
\end{subequations}
Lithium intercalation causes lattice expansion or contraction depending on the active particle chemistry. This chemically driven deformation must be incorporated with the total deformation gradient, $\boldsymbol{F} = \partial\boldsymbol{\varphi}/\partial\boldsymbol{X} = \boldsymbol{1} + \partial\boldsymbol{u}/\partial\boldsymbol{X}$, where $\boldsymbol{X}$ is the reference position,  $\boldsymbol{\varphi}$ is the deformation and $\boldsymbol{u}$ is the displacement field. The multiplicative decomposition $\boldsymbol{F} = \boldsymbol{F}^\text{e}\boldsymbol{F}^\text{c}$ achieves this via the elastic and chemical parts of the deformation gradient $\boldsymbol{F}^\text{e}$ and $\boldsymbol{F}^\text{c}$, respectively. Here, we will consider intercalation strain in the active particles, only, and $\boldsymbol{F}^\text{c}$ will be a function of the Li molar concentration $c_\text{Li}$, which is defined on the deformed configuration of active particles, $\Omega^- = \boldsymbol{\varphi}(\Omega^-_0)$. Cation molar concentrations $c_{\text{Li}^+}$ are defined on the deformed configuration of the electrolyte $\Omega^+ = \boldsymbol{\varphi}(\Omega^+_0)$. The corresponding fluxes are $\boldsymbol{j}_\text{Li}$ and $\boldsymbol{j}_{\text{Li}^+}$ on $\Omega^-$ and $\Omega^+$, respectively. For consistency with the preceding treatment, the electric potential will be denoted by $\phi_\text{e}$ in the solid electrolyte $\Omega^+$, and $\phi_\text{p}$ in the active particle subdomains $\Omega^-$, respectively. In what follows, the electro-chemical governing equations will be posed in the deformed configurations $\Omega^\pm$ with interfaces $\Gamma^\text{a},\Gamma^\text{c}$, while those for mechanics will be in the reference configuration $\Omega_0$ with interfaces $\Gamma^\text{a}_0,\Gamma^\text{c}_0$. Transformations will be invoked only as needed, and not broadly.



\subsubsection{Mass and charge transport}
In the active particles, $\Omega^-$, Li transport reduces to a conservation equation
\begin{equation}
  \frac{\partial c_\text{Li}}{\partial t} + \nabla \cdot \Bj_\text{Li} = 0 
  \WITH 
  \Bj_\text{Li} = -D_\text{Li} \nabla c_\text{Li}\quad \text{in }\Omega^-
  \label{eq:Li-diff-flux}
\end{equation}
where $D_\text{Li}$ is the diffusivity. 

The Li$^+$ cations are also governed by a conservation equation over $\Omega^+$ that has the same form as \eref{eq:Li-diff-flux}:
\begin{subequations}
    \begin{align}
          \frac{\partial c_\text{Li+}}{\partial t} + \nabla \cdot \Bj_{\text{Li}^+} &= 0
  \WITH 
  \Bj_{\text{Li}^+} = -D_{\text{Li}^+} \nabla c_\text{Li+} + \frac{t_+}{F}\Bi^+\quad\text{in }\Omega^+,
  \label{eq:Li+-diff-flux}\\
  \boldsymbol{j}_{\text{Li}^+}\cdot\boldsymbol{n}^+ &= 0,\label{eq:Li+-BC1}\quad\text{on }\partial\Omega^+\backslash\Gamma\\
  -\boldsymbol{i}^+\cdot\boldsymbol{n}^+ &= i_\text{ext}\quad\text{on }\partial\Omega^+\backslash\Gamma\label{eq:Li+-BC2}
    \end{align}
\end{subequations}

where  $D_{\text{Li}^+}$ is the cation diffusivity, $t_+$ is the transference number (the fraction of the total current carried by the cations) and $F$ is the Faraday constant. The current is given by 
\begin{equation}
  \Bi^+ = -\kappa_\text{e}\nabla \phi_\text{e} - \frac{2R\theta\kappa^+}{F}(1-t_+) \nabla \ln c_\text{Li+}\\
\label{eq:Li+-flux-simple}
\end{equation}
where $\kappa_\text{e}$ is the electrolyte's conductivity, $R$ is the universal gas constant and $\theta$ is the temperature. Equation \eref{eq:Li+-flux-simple} corresponds to the general form \cite{ganser2019finite} reduced to a single charged species, dilute in terms of $c_{\text{Li}^+}$.

We turn to the question of boundary conditions for the Li transport equation \eref{eq:Li-diff-flux} over $\Omega^-$. Instead of boundary conditions, interface conditions hold on $\Gamma = \partial\Omega^-$, for Li$^+$ transport, and are discussed below. However, the vanishing flux boundary condition \eref{eq:Li+-BC1} and the current continuity boundary condition \eref{eq:Li+-BC2} hold on $\partial\Omega^+\backslash\Gamma$; this is the external boundary of the electrolyte where it connects to the current collector. 

\subsubsection{Electrostatics}

The electric potential in the active particles is governed by Ohm's law--alternately Gauss' law subject to the electroneutrality condition:
\begin{equation}
  \nabla \cdot \left(-\kappa_\text{p} \nabla \phi_\text{p}\right) = 0\quad \text{in }\Omega^-
  \label{eq:gausslaw-particle}
\end{equation}
where $\kappa_\text{p}$ is the active particle's conductivity. The electric potential in the electrolyte also satisfies the electroneutrality condition:
\begin{subequations}
\begin{align}
      \nabla \cdot \left(-\kappa_\text{e} \nabla \phi_\text{e} - \frac{2 R \theta \kappa_\text{e}}{F}(1-t_+) \nabla \ln c_\text{Li+}\right) &= 0\quad\text{in }\Omega^+\label{eq:gausslaw-electrolyte1}\\
     - \left(-\kappa_\text{e} \nabla \phi_\text{e} - \frac{2 R \theta \kappa_\text{e}}{F}(1-t_+) \nabla \ln c_\text{Li+}\right)\cdot\boldsymbol{n}^+ &= i_\text{ext}\quad\text{on }\partial\Omega^+\label{eq:gausslaw-electrolyte2}
\end{align}
\end{subequations}

As is the case for mass transport over $\Omega^-$, boundary conditions on \eref{eq:gausslaw-particle} are replaced by interface conditions on $\Gamma = \partial\Omega^-$. These will involve $\boldsymbol{j}^-, \boldsymbol{j}^+$ and depend on $\phi_\text{p},\phi_\text{e}$, thus providing the additional condition on $\phi_\text{e}$ needed at $\Gamma$ and coupling the mass/charge transport and electrostatic equations.

\subsubsection{Chemo-mechanics driven by finite intercalation strains} \label{sec:itercalationstrain}

Here, we restrict our model chemistries to those in which intercalation strain arises only in the active particles. The kinematics of intercalation strain are modelled by a multiplicative decomposition of the deformation gradient. This is a treatment that is common to models of chemically induced strain such as from thermal oxidation of silicon \cite{rao2000modelling,Garikipati2001Rao}, lithium intercalation in liquid and solid electrolyte batteries \cite{Wang+Garikipati2017,Wang+Garikipati2018,ganser2019finite,ganser2020stiffer,ganser2021electro,iqbal2021progressive,rezaei2021consistent,rezaei2023cohesive,bistri2021amodeling}, as well as to phenomena of biological growth \cite{garikipati2009kinematics,ambrosi2011perspectives,garikipati2017perspectives}. The multiplicative decomposition is introduced as:
\begin{subequations}
    \begin{align}
        \boldsymbol{F} &= \boldsymbol{F}^\text{e}\boldsymbol{F}^\text{c}\label{eq:fefc}\\
        \boldsymbol{F}^\text{c} &= \left(g(c_\text{Li})\right)^{1/3}\boldsymbol{1}.\label{eq:fc}
    \end{align}
\end{subequations}
where $\boldsymbol{F}^\text{c}$ is the chemical component, which in general introduces incompatibility to the deformation gradient field, and the intercalation function $g(c_\text{Li})$ is specified below. Compatibility is restored by the elastic component of the deformation gradient, $\boldsymbol{F}^\text{e}$, which is also incompatible in general. The multiplicative decomposition is local by definition, as suggested by its illustration for a neighborhood in Figure \ref{fig:deformation-map}. Generally non-uniform fields $c_\text{Li}$ introduce inhomogeneous $\boldsymbol{F}^\text{c}$. For chemo-mechanics of hyperelastic solids, the stresses depend on $\boldsymbol{F}^\text{e}$ in an objective manner. The strain energy density function is a component of the free energy density: $\psi_\text{m}(\boldsymbol{F}^\text{e}) = \widehat{\psi}_\text{m}(\boldsymbol{E}^\text{e})$ for the elastic Green-Lagrange strain tensor $\boldsymbol{E}^\text{e} = \frac{1}{2}(\boldsymbol{F}^{\text{e}^\text{T}}\boldsymbol{F}^\text{e} - \boldsymbol{1})$. Invoking the right Cauchy-Green tensor $\boldsymbol{C} = \boldsymbol{F}^\text{T}\boldsymbol{F}$ we also have $\boldsymbol{E}^\text{e} = \frac{1}{2}(\boldsymbol{F}^{\text{c}^\text{-T}}\boldsymbol{C}\boldsymbol{F}^{\text{c}^{-1}} - \boldsymbol{1})$. We therefore write $\psi = \widehat{\psi}_\text{m}(\boldsymbol{C},c_\text{Li})$, making the chemo-mechanical coupling clear in the strain energy density.

The first Piola-Kirchhoff stress is $\boldsymbol{P} = \boldsymbol{F}^\text{e}(\partial\widehat{\psi}_\text{m}/\partial\boldsymbol{E}^\text{e})\boldsymbol{F}^{\text{c}^{-1}}$ and satisfies the equilibrium equation with boundary conditions
\begin{subequations}
\begin{align}
        \mathrm{DIV}[\boldsymbol{P}] &= \boldsymbol{0}\quad\text{in }\Omega_0\label{eq:stresseq}\\
        \boldsymbol{u} &= \bar{\boldsymbol{u}}\quad\text{on }\partial\Omega_{0u}\label{eq:dirbcu}\\
        \boldsymbol{P}\boldsymbol{N} &= \boldsymbol{T}\quad\text{on }\partial\Omega_{0P}\label{eq:neumannbcP}
\end{align}
\end{subequations}
Here, Dirichlet boundary conditions on $\boldsymbol{u}$ are $\bar{\boldsymbol{u}} = \boldsymbol{0}$ applied where the current collector connects to the solid electrolyte or an active particle. Neumann boundary conditions $\boldsymbol{T} = \boldsymbol{0}$ are applied on lateral surfaces of the configurations presented in Section \ref{sec:simulation}. The above description holds in the absence of interfaces, the case of interest, which will be developed in Section \ref{sec:mechintfc}.

\subsection{Electro-chemo-mechanical coupling in the free energy}
\label{sec:freeenergy}

The free energy density defined on $\Omega_0$ has contributions from the electrostatic displacement and the Li/Li$^+$ concentrations in addition to its mechanical component from the strains. The coupled electro-chemo-mechanics is a consequence and is reflected in governing equations and constitutive relations. We begin by writing
\begin{equation}
    \psi = \psi_\text{e} + \psi_\text{c} + \psi_\text{m}\label{eq:psi}
\end{equation}
for electrostatic and chemical components $\psi_\text{e}$ and $\psi_\text{c}$, respectively. 

For the electrostatic component we write
\begin{equation}
    \psi_\text{e} = \widehat{\psi}_\text{e}(\mathbb{D}) = \frac{1}{2\epsilon_0}\mathbb{D}\cdot\boldsymbol{\epsilon}_r^{\pm^{-1}}\mathbb{D}
    \label{eq:psie}
\end{equation}
where $\epsilon_0$ is the permittivity of vacuum, $\boldsymbol{\epsilon}_r^\pm$ is the relative permittivity of the active particle/electrolyte and $\mathbb{D}$ is the electric displacement vector on $\Omega_0$. The electric field on $\Omega_0$ is 
\begin{equation}
    \mathbb{E} = \frac{\partial\psi_\text{e}}{\partial\mathbb{D}}.
    \label{eq:eleccfield}
\end{equation}
The electric field on $\Omega$ is $\boldsymbol{e} = \boldsymbol{F}^{-\text{T}}\mathbb{E}$ and satisfies $\boldsymbol{e}_{\text{e}/\text{p}} = -\nabla\phi_{\text{e}/\text{p}}$. Here we solve the governing electrostatics directly in terms of $\phi_\pm$ in \eref{eq:gausslaw-particle} and \eref{eq:gausslaw-electrolyte1} by defining the current:
\begin{subequations}
    \begin{align}
        \boldsymbol{i}^- &= -\kappa_\text{p}\nabla\phi_\text{p},\quad -\nabla\cdot\boldsymbol{i}^- = 0\label{eq:freeenergyactivepartgausslaw}\\
        \boldsymbol{i}^+ &= -\kappa_\text{e}\nabla \phi_\text{e} - \frac{2R\theta\kappa_\text{e}}{F}(1-t_+) \nabla \ln c_\text{Li+},\quad -\nabla\cdot\boldsymbol{i}^+ = 0\label{eq:freeenergyelectrolytegausslaw}
    \end{align}
\end{subequations}
Equations \eref{eq:psie}-\eref{eq:freeenergyelectrolytegausslaw} relate the electrostatic governing equations to the electrostatic free energy density. Furthermore, \eref{eq:freeenergyactivepartgausslaw}, \eref{eq:freeenergyelectrolytegausslaw} can be obtained as the Euler-Lagrange equations arising from the extremization of $\widehat{\psi}_\text{e}(\mathbb{D})$.

The chemical component of the free energy is
\begin{equation}
    \psi_\text{c} = \widehat{\psi}_\text{c}(c_\text{Li},c_{\text{Li}^+}) = {\text{det}[\boldsymbol{F}]}c_\text{Li}\mu_\text{Li}^\text{ref} 
    + R\theta \int\limits_{c_\text{Li}^\text{ref}}^{c_\text{Li}} \ln\left(\frac{c_\text{Li}}{c_\text{Li}^\text{ref}} \right)\mathrm{d}c_\text{Li} + {\text{det}[\boldsymbol{F}]}c_{\text{Li}^+}\mu_{\text{Li}^+}^\text{ref} + R\theta \int\limits_{c_{\text{Li}^+}^\text{ref}}^{c_{\text{Li}^+}} \ln\left(\frac{c_{\text{Li}^+}}{c_{\text{Li}^+}^\text{ref}} \right)\mathrm{d}c_{\text{Li}^+}
    \label{eq:psic}
\end{equation}
where $\mu_\text{Li}^\text{ref}$ and $\mu_{\text{Li}^+}^\text{ref}$ are molar reference chemical potentials that are independent of $c_\text{Li},c_{\text{Li}^+},\phi_\pm,\boldsymbol{F}$. The chemical potentials of Li and Li$^+$ have contributions $\mu_{c_\text{Li}} = \partial\psi_\text{c}/\partial c_\text{Li}$ and $\mu_{c_{\text{Li}^+}} = \partial\psi_\text{c}/\partial c_{\text{Li}^+}$:
\begin{subequations}
    \begin{align}
        \mu_{c_\text{Li}} &= {\text{det}[\boldsymbol{F}]}\mu_\text{Li}^\text{ref} + R\theta\ln\left(\frac{c_\text{Li}}{c_\text{Li}^\text{ref}}\right)\label{eq:muc}\\
        \mu_{c_{\text{Li}^+}} &= {\text{det}[\boldsymbol{F}]}\mu_{\text{Li}^+}^\text{ref} + R\theta \ln\left(\frac{c_{\text{Li}^+}}{c_{\text{Li}^+}^\text{ref}} \right) \label{eq:muc+}
    \end{align}
\end{subequations}

of which, the first sub-equation yields the form of the Fickian diffusion term in \eref{eq:Li-diff-flux} for the dilute limit of $c_\text{Li}$. For the solid electrolyte, we introduce the transference number, $t_+$, representing the fraction of current carried by Li$^+$ in the absence of diffusion. Some algebra brings us to the form of the Li$^+$ flux in \eref{eq:Li+-diff-flux}, which is in agreement with the treatment in \cite{ganser2019finite}. 

The forms of the equations \eref{eq:Li-diff-flux} and \eref{eq:Li+-diff-flux} are obtained from ``purely electro-chemical" contributions to $\mu_{c_\text{Li}}$ and $\mu_{c_{\text{Li}^+}}$. Chemo-mechanical coupling furnishes a further driving force, whose form depends on the strain energy density function. Here, we use the St. Venant-Kirchhoff model for the solid electrolyte and active particles, with Lam\'{e} constant $\lambda_{\text{e}/\text{p}}$ and shear modulus $G_{\text{e}/\text{p}}$
\begin{equation*}
    \widehat{\psi}_\text{m}(\boldsymbol{C},c_\text{Li}) = \frac{1}{2}\lambda_{\text{e}/\text{p}}\left(\text{tr}\left[\boldsymbol{E}^\text{e}\right]\right)^2 + G_{\text{e}/\text{p}}\boldsymbol{E}^\text{e}\colon\boldsymbol{E}^\text{e}
\end{equation*}
which on accounting for the elasto-chemo decomposition of $\boldsymbol{F}$ in \eref{eq:fefc} and \eref{eq:fc} yields the following form in the active particles:
\begin{equation}
    \widehat{\psi}^-_\text{m}(\boldsymbol{C},c_\text{Li}) = \frac{1}{2}\lambda_\text{p}\left( \frac{\left(g(c_\text{Li})\right)^{-2/3}}{2}\text{tr}\left[\boldsymbol{C}\right] - \frac{3}{2}\right)^2 +  G_\text{p}\left[\frac{1}{2}(\left(g(c_\text{Li})\right)^{-2/3}\boldsymbol{C} - \boldsymbol{1}) \right]\colon\left[\frac{1}{2}(\left(g(c_\text{Li})\right)^{-2/3}\boldsymbol{C} - \boldsymbol{1}) \right].
    \label{eq:SVK}
\end{equation}

The total chemical potential in the active particles therefore is
\begin{equation}
    \mu^-_{c_\text{Li}}(\boldsymbol{C},c_\text{Li}) = {\text{det}[\boldsymbol{F}]}\mu_\text{Li}^\text{ref} + R\theta\ln\left(\frac{c_\text{Li}}{c_\text{Li}^\text{ref}}\right) + (\lambda_{\text{e}/\text{p}} + \frac{2 G_{\text{e}/\text{p}}}{3})\left(  \frac{\left(g(c_\text{Li})\right)^{-2/3}}{2}\text{tr}\left[\boldsymbol{C}\right] - \frac{3}{2}\right) \left(-\frac{1}{3}\left(g(c_\text{Li})\right)^{-4/3} \text{tr}\left[\boldsymbol{C}\right]\right),\label{eq:muctot}
\end{equation}
making explicit the chemo-mechanical coupling in terms of $\boldsymbol{C}$ and $c_\text{Li}$. A more transparent form, which can be arrived at via elementary tensor calculus and by mapping of stress measures between $\Omega_0$ and $\Omega$ is
\begin{equation}
    \mu^-_{c_\text{Li}}(\boldsymbol{C},c_\text{Li}) = {\text{det}[\boldsymbol{F}]}\mu_\text{Li}^\text{ref} + R\theta\ln\left(\frac{c_\text{Li}}{c_\text{Li}^\text{ref}}\right) - \frac{1}{3}\text{det}[\boldsymbol{F}]\left(g(c_\text{Li})\right)^{-2/3}\text{tr}[\boldsymbol{\sigma}],\label{eq:muctot1}
\end{equation}
where  $\boldsymbol{\sigma}$ is the Cauchy stress.

The Li flux in active particles is

\begin{equation*}
    \boldsymbol{j}_\text{Li} = -M_\text{Li}c_\text{Li}(1-\frac{c_\text{Li}}{c_\text{Li}^\text{max}})\nabla\mu^-_{c_\text{Li}}
\end{equation*}

where $M_\text{Li}c_\text{Li}(1-c_\text{Li}/c_\text{Li}^\text{max})$ is the mobility. For dilute conditions, $c_\text{Li} \ll c_\text{Li}^\text{max}$, we have
\begin{equation}
    \boldsymbol{j}_\text{Li} = -M_\text{Li} R\theta \nabla c_\text{Li} -M_\text{Li}c_\text{Li}\nabla\left( \text{det}[\boldsymbol{F}]\left(g(c_\text{Li})\right)^{-2/3}\frac{1}{3}\text{tr}[\boldsymbol{\sigma}]\right),
    \label{eq:chemomechflux}
\end{equation}
where $M_\text{Li} R\theta = D_\text{Li}$.  Noting that $\frac{1}{3}\text{tr}[\boldsymbol{\sigma}$ is the hydrostatic stress, the traditional form of pressure gradient-driven mass transport is seen in the flux relation. This constitutive model satisfies the dissipation inequality; see Ganser et al. \cite{ganser2019finite} for a comprehensive treatment, which we do not revisit here.

\section{Continuum treatment of interfaces }\label{sec:interface}

The treatment of interfaces in continuum physics is natural, especially when approached in the integral form of the governing equations \cite{truesdell1960classical,truesdell2004non}. We lay out the equations in weak form beginning with mass balance.

\subsection{Mass balance for a scalar field in the presence of an interface}

For mass balance of a single component whose concentration is denoted by $c(\boldsymbol{x},t)$ the strong form in the presence of an interface $\Gamma\subset\Omega$ is:
\begin{equation}
  \begin{aligned}
    \frac{\partial c}{\partial t} + \nabla \cdot \Bj = 0 \quad & \text{in} \quad \Omega\backslash \Gamma, \\
    c_0 (\boldsymbol{x},0) = \bar{c}_0 (\boldsymbol{x})  \quad & \text{on} \quad  \Omega\backslash \Gamma, \\
    c (\boldsymbol{x},t) = \bar{c} (\boldsymbol{x},t)  \quad & \text{on} \quad  \partial\Omega^{c}\times [0,T], \\
    -\boldsymbol{j}\cdot\boldsymbol{n} = \bar{j} (\boldsymbol{x},t)  \quad & \text{on} \quad  \partial\Omega^j\times [0,T].
  \end{aligned}
  \label{eq:governing-eq-C}
\end{equation}
Additionally, we allow for interface reactions on  $\Gamma$ with a rate $R (\boldsymbol{x}, c^+, c^-)$. 

\subsubsection{The weak form of scalar transport equations with an interface}
\label{sec:weakformgentransp}
The weak form of mass balance equation is written as: Given $w \in \mathcal{V}$ find $c \in \mathcal{S}$  such that

\begin{equation}
    \begin{aligned}
        \int_{\Omega}  w \frac{\partial c}{\partial t}dV
        & = \int_{\Omega} \nabla w \cdot \Bj dV + \int_{\partial \Omega} w \bar{j} dS - \int_{\Gamma} w \llbracket\Bj \cdot \Bn\rrbracket \mathrm{d}S  \\
    \end{aligned}
    \label{eq:weak-form-diffusion-omega}
\end{equation}
where the flux discontinuity is $\llbracket\boldsymbol{j}\cdot\boldsymbol{n}\rrbracket = \boldsymbol{j}^+\cdot\boldsymbol{n}^+ + \boldsymbol{j}^-\cdot\boldsymbol{n}^-$. We recall the important technical point that the integral over $\Omega$ is strictly  over $\Omega^+\cup \Omega^- = \Omega\backslash\Gamma$. However for regular (non-singular) integrands, the integrals over either domain are equal since $\Gamma$ is a set of zero measure in $\mathbb{R}^3$. Here, we are interested in interface reactions of the form introduced above, that drive the flux discontinuity:
\begin{equation}
    \llbracket\Bj \cdot \Bn\rrbracket = R(\boldsymbol{x}, c^+, c^-).
    \label{eq:flux-jump-along-gamma}
\end{equation}
The flux continuity condition across $\Gamma$ is $ \llbracket\Bj \cdot \Bn\rrbracket = 0$ for $R = 0$
Since Equations \eref{eq:weak-form-diffusion-omega} and \eref{eq:flux-jump-along-gamma} are restricted to being only in terms of $c$, we have not introduced additional fields in the functional dependence of $R$. An extension of that functional dependence is natural for electro-chemical coupling and is considered below.  We first rewrite \eref{eq:weak-form-diffusion-omega} as weak forms over $\Omega^+$ and $\Omega^-$, specifying that the transported species in an active particle, $\Omega^-$, is Li with concentration $c_{\text{Li}}$, and in the solid electrolyte, $\Omega^+$, it is Li$^+$ with concentration $c_{\text{Li}^+}$. We have \cite{Wang+Garikipati2018}:

\begin{equation}
    \begin{aligned}
        \int_{\Omega^-}  w \frac{\partial c_{\text{Li}}}{\partial t}dV
        & = \int_{\Omega^-} \nabla w \cdot \Bj_{\text{Li}} dV - \int_{\partial \Omega^-} w \boldsymbol{j}_\text{Li}\cdot\boldsymbol{n} dS - \int_{\Gamma} w\Bj_{\text{Li}}\cdot \Bn^{+}dS  \\
        \int_{\Omega^+}  w \frac{\partial c_{\text{Li}^+}}{\partial t}dV
        & = \int_{\Omega^+} \nabla w \cdot \Bj_{\text{Li}^+} dV - \int_{\partial \Omega^+} w \boldsymbol{j}_{\text{Li}^+}\cdot\boldsymbol{n} dS - \int_{\Gamma} w\Bj_{\text{Li}^+} \cdot \Bn^{-}dS,
    \end{aligned}
    \label{eq:weak-form-diffusion-two-region-revised}
\end{equation}
where the fluxes $\Bj_\text{Li}$  and $\Bj_{\text{Li}^+}$ satisfy \eref{eq:Li-diff-flux}, \eref{eq:Li+-diff-flux} and \eref{eq:Li+-flux-simple}. We similarly rewrite the interface reaction equation: 
\begin{equation}
\Bj_{\text{Li}^+}\cdot \Bn^{+} + \Bj_\text{Li} \cdot \Bn^{-} = R(\boldsymbol{x}, c^+, c^-)\;\text{on } \Gamma.
    \label{eq:weak-form-diffusion-omega-jump-w}
\end{equation}

\subsubsection{The embedded interface treatment}
\label{sec:embeddedintfc}
\begin{figure}[h!]
    \centering
    \includegraphics[width=0.9\textwidth]{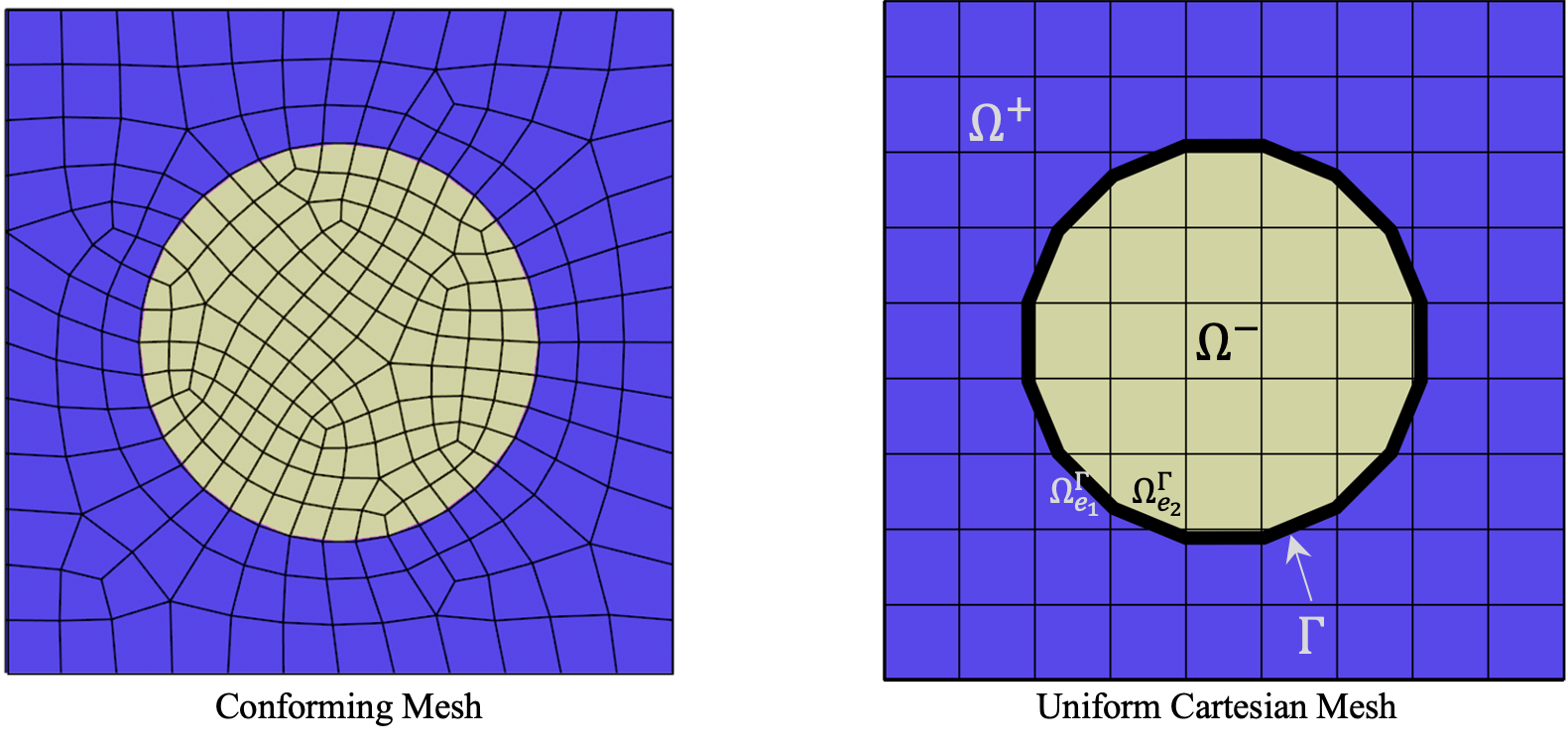}
    \caption{Left: conforming finite element mesh resolving an idealized two-dimensional particle geometry; right: uniform Cartesian mesh, with the interface $\Gamma$ passing through the elements $\Omega^\Gamma_{e_1}, \Omega^\Gamma_{e_2},\dots$.}
    \label{fig:ill-composite-ss-electrode}
\end{figure}

The mathematical treatment of the interface is central to this work. We use finite element methods, and Figure \ref{fig:ill-composite-ss-electrode} illustrates a feature-conforming finite element mesh for circular particles. While following the interface is relatively easy for simple geometries, arbitrary particle shapes lead to increasingly complicated meshes.  Here, rather than adopt conforming meshes to resolve active particles geometries, we represent the particle-electrolyte interface, $\Gamma$ as the zero levelset of a scalar function, $\eta(\boldsymbol{x})$ against a background Cartesian mesh, as also illustrated in Figure \ref{fig:ill-composite-ss-electrode}. Thus, the following will be implied whenever $\Gamma$ is referred to, especially in the mesh-based context:
\begin{equation}
    \boldsymbol{x} \in \Gamma,\quad\forall \boldsymbol{x}\text{ s.t. }\eta(\boldsymbol{x}) = 0
    \label{eq:defGamma}
\end{equation}
An element, $\Omega_e$ that is intersected by a subset of the interface $\Gamma_e \subset \Gamma$ has subsets $\Omega_e^+$ and $\Omega_e^-$ such that $\overline{\Omega_e} = \overline{\Omega_e^+\cup \Omega_e^- \cup \Gamma}$. Furthermore the fluxes local to this element admit jumps
\begin{subequations}
    \begin{align}
    \llbracket \boldsymbol{j}_{\text{Li}^+}\rrbracket &= \boldsymbol{j}_{\text{Li}^+}^+\cdot\boldsymbol{n}^+ + \boldsymbol{j}_{\text{Li}^+}^-\cdot\boldsymbol{n}^-\; \text{on } \Gamma, \quad\text{s.t. }\boldsymbol{j}_{\text{Li}^+}^- = \boldsymbol{0}\;\text{in }\Omega_e^-,\label{eq:flux-jumps-Li+}\\  
    \llbracket \boldsymbol{j}_{\text{Li}}\rrbracket &= \boldsymbol{j}_{\text{Li}}^+\cdot\boldsymbol{n}^+ + \boldsymbol{j}_{\text{Li}}^-\cdot\boldsymbol{n}^-\; \text{on } \Gamma, \quad\text{s.t }\boldsymbol{j}_{\text{Li}}^+ = \boldsymbol{0}\;\text{in }\Omega_e^+.
    \label{eq:flux-jumps-Li}
\end{align}
\end{subequations}

The above equations mean that the flux of Li$^+$ vanishes in the particle and flux of Li vanishes in the electrolyte. However, the respective quantities $\boldsymbol{j}_{\text{Li}^+}^-$ and $\boldsymbol{j}_{\text{Li}}^+$ must be represented in $\Omega_e^-$ and $\Omega_e^+$. This also implies that the concentrations $c^+$ and $c^-$ must be represented in $\Omega_e^-$ and $\Omega_e^+$, respectively, and that these fields also suffer discontinuities
\begin{subequations}
    \begin{align}
    \llbracket c^+_{\text{Li}^+}\rrbracket &= c_{\text{Li}^+} - c_{\text{Li}^+}^-\; \text{on } \Gamma, \quad\text{s.t. }c_{\text{Li}^+}^- = 0\;\text{in }\Omega_e^-,\label{eq:concjumpsLi+}\\  \llbracket c_{\text{Li}}\rrbracket &= c_{\text{Li}}^+ - c_{\text{Li}}^-\; \text{on } \Gamma, \quad\text{s.t. }c_{\text{Li}}^+ = 0\;\text{in }\Omega_e^+.\label{eq:concjumpsLi}
\end{align}
\end{subequations}

The combination of \eref{eq:weak-form-diffusion-omega-jump-w}, with jump conditions \eref{eq:flux-jumps-Li+} and \eref{eq:flux-jumps-Li} is to be imposed on $\Gamma \cap\Omega_e$. For brevity we introduce the notation $\Omega_e^\Gamma$ for an element that satisfies $\Gamma \cap\Omega_e \neq \emptyset$

In this work we treat the above discontinuous fields by the strong discontinuity approach, which has been used in reaction-transport problems previously \cite{rao2000modelling}, \cite{Garikipati2001Rao}. Originally introduced for displacement discontinuities in inelastic solids \cite{armero1996analysis,oliver1996modelling,regueiro2001plane,oliver2004continuum,linder2007finite,armero2008new}, here, we apply it more widely to the reaction-transport, nonlinear elastic fracture as well as electrostatic problems--that is, to all of the electro-chemo-mechanics of solid state batteries. Starting with the reaction-transport problem we consider the treatment of a concentration field, $c$, admitting a discontinuity: 

\begin{equation}
    c  = \bar{c} + \llbracket c\rrbracket,\quad \text{where }\llbracket c\rrbracket = H_\Gamma \xi
    \label{eq:discont-c0}
\end{equation}
with $\bar{c}$ being the continuous component, $H_\Gamma$ the Heaviside on $\Gamma$ defined by
\begin{equation}
    H_\Gamma(\Bx) =
    \begin{cases}
        1  \quad \text{if} ~ \Bx \in ~\Omega^+ \\
        0  \quad \text{if} ~ \Bx \in ~\Omega^- \\
    \end{cases}
\end{equation}
and $\xi$ a scalar. Equation \eref{eq:discont-c0} can be rewritten over $\Omega_e^\Gamma$ as
\begin{equation}
    c  = \tilde{c} + M_\Gamma\xi,\quad\text{in }\Omega_e^\Gamma
    \label{eq:discont-c1}
\end{equation}
where $\tilde{c}$ is the representation from continuous basis functions and $M_\Gamma$ satisfies
\begin{equation}
    M_\Gamma(\Bx) = H_\Gamma(\Bx) - \chi(\Bx).
    \label{eq:discont-M}
\end{equation}
This accounts for the difference between the true discontinuity $H_\Gamma(\boldsymbol{x})$ and its approximation in the continuous basis $\chi(\boldsymbol{x})$. The latter function has the property
\begin{equation}
    \chi(\Bx) =
    \begin{cases}
        1  \quad \text{if} ~ \Bx \in ~\Omega^+\backslash \Omega_e^\Gamma \\
        0  \quad \text{if} ~ \Bx \in ~\Omega^-\backslash \Omega^\Gamma_e, \\
    \end{cases}
\end{equation}
which enforces $M_\Gamma(\Bx) = 0$ in elements that do not intersect $\Gamma$, and is only non-zero in elements intersecting $\Gamma$.

\begin{figure}[h!]
    \centering
    \includegraphics[width=1.0\linewidth]{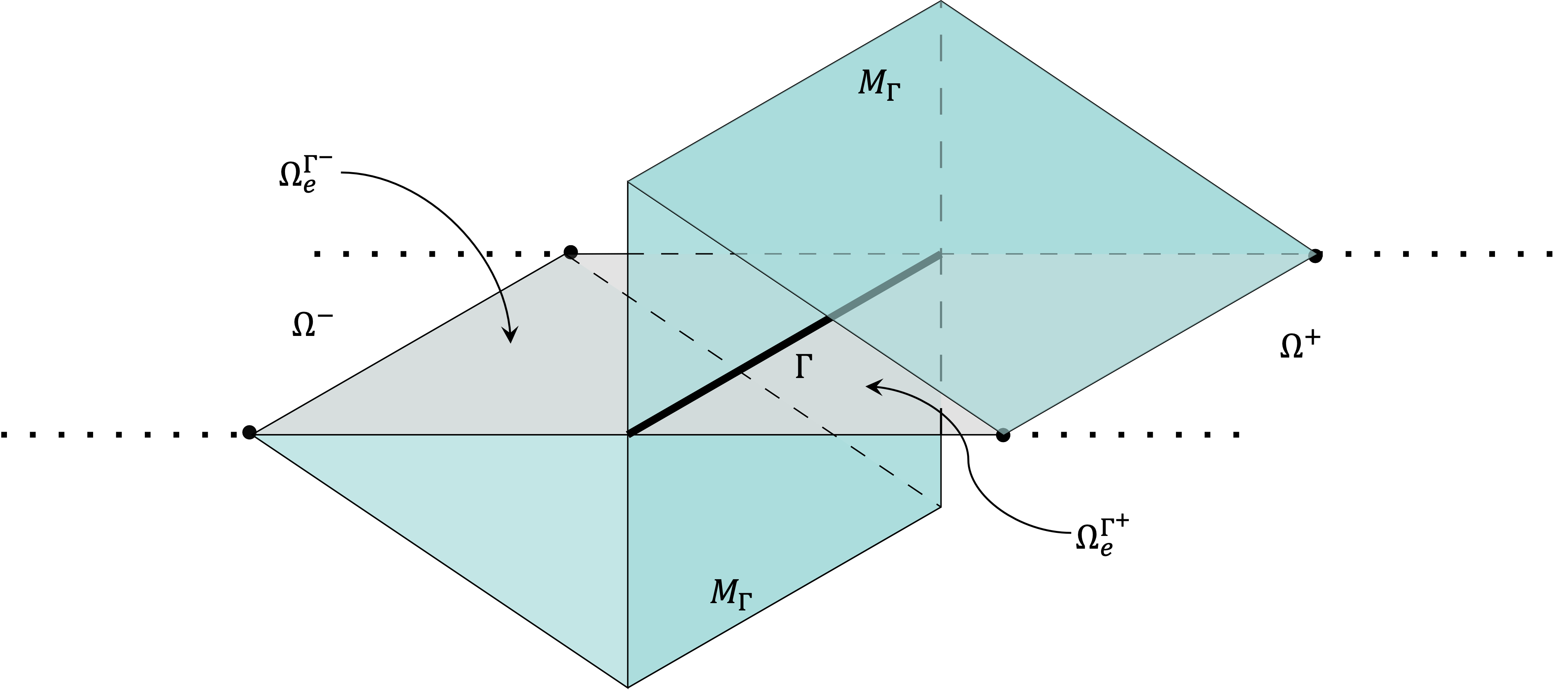}
    \caption{Illustration of the discontinuous basis function, $M_\Gamma$. 
    }
    \label{fig:element-jumps}
\end{figure}

In the simplest case, which we adopt here, $\llbracket c(\boldsymbol{x})\rrbracket$ does not vary along $\Gamma$ within $\Omega_e^\Gamma$. See Fig. \ref{fig:element-jumps} for an illustration of this case. The treatment here follows the non-conforming finite element approach where variations along $\Gamma$ are obtained by $\xi$ taking on different values in adjacent elements $\Omega_{e_1}^\Gamma$ and $\Omega_{e_2}^\Gamma$ both of which have non-empty intersections with $\Gamma$. Then restricting to Lagrange polynomial basis functions $N^A(\boldsymbol{x})$ that have the Kronecker-delta property $N^A(\boldsymbol{x}_B) = \delta_{AB}$ at finite element nodes placed at $\boldsymbol{x}_B \in \Omega_e^\Gamma$ a convenient representation for $\chi(\boldsymbol{x})$ is
\begin{equation}
    \chi(\boldsymbol{x}) = \sum\limits_{A,\;\mathrm{s.t.} \boldsymbol{x}_A\in\Omega_e^{\Gamma^+}} N^A(\boldsymbol{x}).
    \label{eq:chi}
\end{equation}

where $\Omega_e^{\Gamma^+}$ is the subdomain into which $\boldsymbol{n}^+$ points. 

\subsubsection{The weak forms for Li transport with reactions across a particle-electrolyte interface}
\label{sec:weakformtranspreac}
The treatment in terms of weak forms in Section \ref{sec:weakformgentransp} is made more specific for Li$^+$ transport through the electrolyte, reaction with electrons $e^-$ at the particle-electrolyte interface and transport of Li through the particle:

\begin{subequations}
    \begin{align}
                \int_{\Omega^-}  w \frac{\partial c_{\text{Li}}}{\partial t}dV
        & = \int_{\Omega^-} \nabla w \cdot \Bj_{\text{Li}} dV - \int_{\partial \Omega^-} w \Bj_{\text{Li}}\cdot \Bn dS - \int_{\Gamma} w\Bj_{\text{Li}}\cdot \Bn^{+}dS  \label{eq:weakmasstransp}\\
        \boldsymbol{j}_\text{Li} &= -D_\text{Li}\nabla c_\text{Li} -M_\text{Li}c_\text{Li}\nabla\left( \text{det}[\boldsymbol{F}]\left(g(c_\text{Li})\right)^{-2/3}\frac{1}{3}\text{tr}[\boldsymbol{\sigma}]\right)\label{eq:Liflux}\\
        \int_{\Omega^+}  w \frac{\partial c_{\text{Li}^+}}{\partial t}dV
        & = \int_{\Omega^+} \nabla w \cdot \Bj_{\text{Li}^+} dV - \int_{\partial \Omega^+} w \Bj_{\text{Li}^+}\cdot \Bn dS - \int_{\Gamma} w\Bj_{\text{Li}^+} \cdot \Bn^{-}dS\label{eq:weakchargetransp}\\
        \boldsymbol{j}_{\text{Li}^+} &= -D_{\text{Li}^+} \nabla c_\text{Li+} - \frac{t_+}{F}\left( \kappa_e \nabla \phi_\text{e} + \frac{2 R \theta \kappa_e}{F}(1-t_+) \nabla \ln c_\text{Li+}\right) \label{eq:Li+flux}
    \end{align}
\end{subequations}

Additionally,  charge transfer kinetics are imposed as an interface condition on $\Gamma$ as detailed in \eref{eq:BV0}. 

\subsubsection{The weak form for the Poisson equation for electric fields with a particle-electrolyte interface} \label{sec:poissoneqphi}

The electric potential in the active particles and the electrolyte also suffers jumps at $\Gamma$ when modelled within the embedded interface treatment of Section \ref{sec:embeddedintfc}:
\begin{align}
    \llbracket \phi_\text{e}\rrbracket &= \phi_\text{e}^+ - \phi_\text{e}^-\; \text{on } \Gamma, \quad\text{s.t. }\phi_\text{e}^- = 0\;\text{in }\Omega_e^-\\  \llbracket \phi_\text{p}\rrbracket &= \phi_\text{p}^+ - \phi_\text{p}^-\; \text{on } \Gamma, \quad\text{s.t. }\phi_\text{p}^+ = 0\;\text{in }\Omega_e^+.
\end{align}
and gradient conditions
\begin{subequations}
    \begin{align} 
    \llbracket \nabla\phi_{\text{e}}\rrbracket &= \nabla\phi_{\text{e}}^+\cdot\boldsymbol{n}^+ + \nabla\phi_{\text{e}}^-\cdot\boldsymbol{n}^-\; \text{on } \Gamma, \quad\text{s.t }\nabla\phi_{\text{e}}^-\cdot\boldsymbol{n}^- = \boldsymbol{0}\;\text{in }\Omega_e^-,
    \label{eq:jumpsgradphie}\\
    \llbracket \nabla\phi_{\text{p}}\rrbracket &= \nabla\phi_{\text{p}}^+\cdot\boldsymbol{n}^+ + \nabla\phi_{\text{p}}^-\cdot\boldsymbol{n}^-\; \text{on } \Gamma, \quad\text{s.t. }\nabla\phi_{\text{p}}^+\cdot\boldsymbol{n}^+ = \boldsymbol{0}\;\text{in }\Omega_e^+,\label{eq:jumpsgradphip}
\end{align}
\end{subequations}
The discontinuous scalar fields $\phi_\text{p}$ and $\phi_\text{e}$ can be represented by the basis function $M_\Gamma$ over elements $\Omega_e^\Gamma$. The corresponding weak forms are:
\begin{align}
    \int\limits_{\Omega^-}\nabla w\cdot\nabla\phi_\text{p} dV &= 0\label{eq:weakelectroparticle}\\
    \int\limits_{\Omega^+}\kappa_e\nabla w \cdot \left(\nabla\phi_\text{e} + \frac{2R\theta}{F}(1-t_+)\nabla\ln C_{\text{Li}^+}\right) dV &= \int\limits_{\partial\Omega^+}i_\text{ext}\mathrm{d}S,\label{eq:weakelectroelectrolyte}
\end{align}



where the Neumann boundary condition imposes current continuity from \eref{eq:Li+-BC2}. This treatment of  $\phi_\text{p}$ and $\phi_\text{e}$ is combined with that of
 $j_\text{Li}$ and $j_{\text{Li}^+}$ in Section \ref{sec:weakformtranspreac} into interface conditions that are given by Butler-Volmer charge transfer kinetics, now reinterpreted in terms of discontinuous fields
\begin{equation}
            \Bj_{\text{Li}^+}\cdot \Bn^{+} + \Bj_\text{Li} \cdot \Bn^{-} = j_0\left(\text{exp}\left(\frac{\alpha_aF}{R\theta}(\phi_\text{p}^- -\phi_\text{e}^+ -U)\right)-\text{exp}\left(-\frac{\alpha_aF}{R\theta}(\phi_\text{p}^- -\phi_\text{e}^+ -U)\right)\right)\quad\text{on } \Gamma
            \label{eq:BV0}
\end{equation}
\subsection{Finite strain kinematics with a discontinuous displacement vector field}
\label{sec:mechintfc}
The interface between solid electrolytes and active particles is susceptible to fracture, especially with brittle ceramics such LLZO. The equation of mechanical equilibrium translates to traction continuity on $\Gamma$:
\begin{equation}
    \llbracket \boldsymbol{PN}\rrbracket = \boldsymbol{P}^+\boldsymbol{N}^+ + \boldsymbol{P}^-\boldsymbol{N}^- = \boldsymbol{0}\quad\text{on } \Gamma.
\end{equation}
In this work, the treatment of interface fracture is not via the creation of free surfaces, which bear zero traction, but as a sharp interface with degraded traction. The finite separation of the free surfaces is represented by a displacement discontinuity. This treatment of interface fracture and the ensuing mechanics follows the strong discontinuity approach \cite{armero1996analysis,oliver1996modelling,regueiro2001plane,oliver2004continuum,linder2007finite,armero2008new,rudraraju2011theory}. We have the following decomposition of the deformation into continuous and discontinuous components:
\begin{equation}
  \Bvarphi(\BX) = \bar{\Bvarphi} + \llbracket \Bvarphi \rrbracket H_{\Gamma_0}(\BX)
\end{equation}
where $H_{\Gamma_0}(\BX)$ is the Heaviside function with respect to the reference configuration. This leads to the deformation gradient:
\begin{equation}
    \boldsymbol{F} = \bar{\boldsymbol{F}} + \llbracket\boldsymbol{\varphi}\rrbracket\otimes\boldsymbol{N}\delta_{\Gamma_0}
\end{equation}
where $\bar{\boldsymbol{F}}$ is the regular (non-singular) component, $\delta_{\Gamma_0}$ is the one-dimensional Dirac-delta at $\Gamma_0$, $\boldsymbol{N}$ is the normal vector to $\Gamma_0$ in the reference configuration, and the tensor product defines the singular component of $\boldsymbol{F}$. Following the approach in Section \ref{sec:embeddedintfc} as well as in \cite{garikipati2000variational,armero2008new} we first write
\begin{equation}
    \bar{\boldsymbol{\varphi}}^h = \sum\limits_{A} N^A(\boldsymbol{X})\boldsymbol{d}^A
\end{equation}
for basis functions $N^A$, which allows us to rewrite the deformation gradient for elements $\Omega_e^\Gamma$:

\begin{equation}
  \BF^h = \Grad[\bar{\Bvarphi}^h] + \tilde{\BF}^h
\end{equation}
where $\Grad\bar{\boldsymbol{\varphi}}^h = \sum_A \Grad N^A\otimes\boldsymbol{d}$. Here, $\tilde{\boldsymbol{F}}^h$ can be considered either as an enhanced strain as in the original strong discontinuity treatment \cite{armero1996analysis} or a fine scale strain in the variational multiscale setting \cite{garikipati2000variational}. In either case it can be expressed in the form
\begin{equation}
  \tilde{\BF}^h = - \Balpha \otimes \frac{\partial\chi(\boldsymbol{X})}{\partial\boldsymbol{X}} + \Balpha \otimes \BN \delta_{\Gamma_0}.
  \label{eq:enhstrain}
\end{equation}
The enhanced strain and variational multiscale treatments lead to two equations in weak form:
\begin{subequations}
    \begin{align}
        \int\limits_{\Omega_0} \Grad\boldsymbol{w}^h\colon\boldsymbol{P} \mathrm{d}V &= \boldsymbol{0}\label{eq:csweakform}\\
        \sum\limits_{\Omega_{e_1}^\Gamma\cup,\dots\Omega_{e_n}^\Gamma}\int\limits_{\Omega_e^\Gamma} \tilde{\boldsymbol{H}}\colon\boldsymbol{P} \mathrm{d}V &= \boldsymbol{0}\label{eq:fsweakform}
    \end{align}
\end{subequations}
In particular \eref{eq:fsweakform} leads to
\begin{equation}
\sum\limits_{e\in\{e_1,\dots e_n\}}\int\limits_{\Omega_e^\Gamma}\boldsymbol{\beta}\cdot\boldsymbol{P}\frac{\partial\chi(\boldsymbol{X})}{\partial\boldsymbol{X}}\mathrm{d}V = \sum\limits_{e\in\{e_1,\dots e_n\}}\int\limits_{\Gamma_e}\boldsymbol{\beta}\cdot\boldsymbol{PN}\mathrm{d}S
\label{eq:avgtrac0}
\end{equation}
with $\Bbeta$ being the variation associated with the enhanced strain or with the fine scale strain, depending on the treatment and $\Gamma_e = \Gamma\cap\Omega_e^\Gamma$. For $\Bbeta$ uniform over $\Omega_e^\Gamma$, and writing $\boldsymbol{PN} = \boldsymbol{T}_\Gamma$, the traction on $\Gamma$, this leads to 
\begin{equation}
\foo_{e\in\{e_1,\dots e_n\}}\int\limits_{\Omega_e^\Gamma}\boldsymbol{P}\frac{\partial\chi(\boldsymbol{X})}{\partial\boldsymbol{X}}\mathrm{d}V = \foo_{e\in\{e_1,\dots e_n\}}\int\limits_{\Gamma_e}\boldsymbol{T}_\Gamma \mathrm{d}S,
\label{eq:avgtrac}
\end{equation}
where $\barr$ is the finite element assembly opertor. Here we assume the shear component of the traction $\boldsymbol{T}_\Gamma$ to vanish
\begin{equation}
  \BT_\Gamma = T_{\Gamma N} \BN
  \label{eq:normaltrac0}
\end{equation}
and $\boldsymbol{\alpha} = \xi_N\boldsymbol{N}$ in \eref{eq:enhstrain}, that is the crack opening on $\Gamma$ is only in the normal direction. For the traction-separation relationship we use  a linear law \cite{armero1996analysis,garikipati2002variational,linder2007finite,rudraraju2011theory}:
\begin{equation}
  \begin{aligned}
    T_{\Gamma N} = \text{max}\left\{ 0, f_t - K \xi_N \right\},
  \end{aligned}
\end{equation}
where $K$ is the softening modulus and $f_t$ is the threshold of the maximum stress.


\subsection{Solution of finite element equations}
\label{sec:fesoln}

The weak forms for mass transport \eref{eq:weakmasstransp}, charge transport \eref{eq:weakchargetransp}, electrostatics \eref{eq:weakelectroparticle} and \eref{eq:weakelectroelectrolyte}, and mechanics \eref{eq:csweakform}, \eref{eq:fsweakform} involve discontinuous fields $\llbracket c_\text{Li}\rrbracket$,  $\llbracket c_{\text{Li}^+}\rrbracket$, $\llbracket \phi_\text{p}\rrbracket$ and $\llbracket \phi_\text{e}\rrbracket$, and $\llbracket \boldsymbol{\varphi}\rrbracket$, respectively. These fields correspond to scalar unknowns denoted by $\xi$ in \eref{eq:discont-c1} for $\llbracket c_\text{Li}\rrbracket$,  $\llbracket c_{\text{Li}^+}\rrbracket$, $\llbracket \phi_\text{p}\rrbracket$ and $\llbracket \phi_\text{e}\rrbracket$ and $\boldsymbol{\alpha} = \xi_N\boldsymbol{N}$ in \eref{eq:enhstrain}. The corresponding weak forms are implemented as finite element residuals
\begin{equation}
    \left\{
    \begin{array}{c}
         \bar{\boldsymbol{R}}_\text{Li} \\
         \bar{\boldsymbol{R}}_{\text{Li}^+} \\
         \bar{\boldsymbol{R}}_\text{p} \\
         \bar{\boldsymbol{R}}_\text{e} \\
         \bar{\boldsymbol{R}}_\varphi 
    \end{array} \right\}= \boldsymbol{0}
    \label{eq:residual0}
\end{equation}
where the respective residual vectors are $\bar{\boldsymbol{R}}_\text{Li},\dots,\bar{\boldsymbol{R}}_\varphi$. The local element residuals for element $e$ have the form
\begin{equation}
    \left\{
    \begin{array}{c}
         \bar{\boldsymbol{R}}^e  \\
          \bar{\boldsymbol{R}}^e_\xi
    \end{array}\right\} = \boldsymbol{0}.
    \label{eq:residual1}
\end{equation}
Of these components, $\bar{\boldsymbol{R}}^e_\xi = \boldsymbol{0}$ represents the contribution imposing the conditions in \eref{eq:flux-jumps-Li+}, \eref{eq:flux-jumps-Li}, \eref{eq:jumpsgradphie}, \eref{eq:jumpsgradphip} and \eref{eq:fsweakform}, respectively. The corresponding finite element degrees of freedom $\xi$ and $\boldsymbol{\alpha}$ are local to the element $e$ and therefore so are their respective variations. Therefore, $\bar{\boldsymbol{R}}^e_\xi = \boldsymbol{0}$ can be solved locally without assembly into the global residual. In our implementation these nonlinear finite element equations are solved by generating the corresponding jacobians via automatic differentiation using the \texttt{Sacado} library of the \texttt{TriLinos} project \cite{heroux2012new}. The jacobian submatrices are extracted from the automatic differentiation implementation and used to iteratively update the jump degrees of freedom, $\xi$ and $\boldsymbol{\alpha}$ locally at the element level. These local iterates yield updated residuals $ \boldsymbol{R}_\text{Li},  \boldsymbol{R}_{\text{Li}^+}, \boldsymbol{R}_\text{p}, \boldsymbol{R}_\text{e},  \boldsymbol{R}_\varphi$ and jacobian submatrices that are reassembled into a global residual represented as
\begin{equation}
    \boldsymbol{R}(\boldsymbol{d}) =  \left\{ \begin{array}{c}
        \boldsymbol{R}_\text{Li} \\
         \boldsymbol{R}_{\text{Li}^+} \\
         \boldsymbol{R}_\text{p} \\
         \boldsymbol{R}_\text{e} \\
         \boldsymbol{R}_\varphi 
    \end{array} \right\}(\boldsymbol{d}) 
    \label{eq:residual2}
\end{equation}
for a global degrees of freedom vector $\boldsymbol{d}$ that includes nodal values corresponding to all the fields: $c_\text{Li}, c_{\text{Li}^+}, \phi_\text{p}, \phi_\text{e}, \boldsymbol{u}$  and solved as 
\begin{equation}
    \boldsymbol{R}(\boldsymbol{d}) + \frac{\partial\boldsymbol{R}}{\partial\boldsymbol{d}}\delta\boldsymbol{d} = \boldsymbol{0}\label{eq:residualsoln}
\end{equation}
where $\partial\boldsymbol{R}/\partial\boldsymbol{d}$ is computed by automatic differentiation.

\section{Coupling between electrochemistry and nonlinear mechanics}
\label{sec:electrochemmech}

\subsection{Degradation of interface charge transfer kinetics with crack opening}

\begin{figure}[h!]
  \centering
  \includegraphics[width=0.7\linewidth]{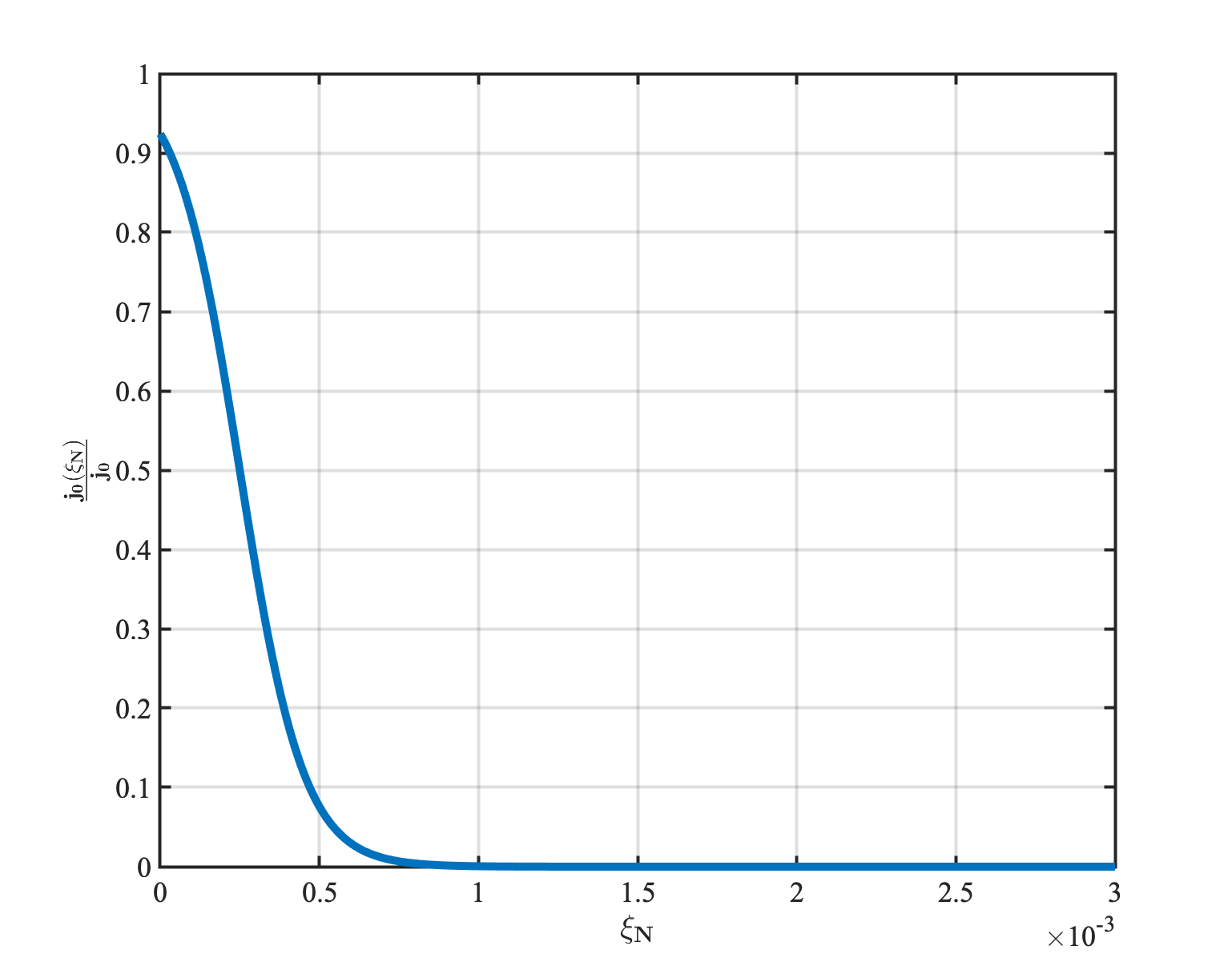}
  \caption{The sigmoid degradation function of charge transfer kinetics with $l = 0.0001$ $\mu$m and $\xi_\text{max} = 0.0005$ $\mu$m.}
  \label{fig:sigmoid}
\end{figure}

The normal crack opening displacement creates a loss of contact between the electrolyte and active particle at their interface, $\Gamma$. The degradation of charge transfer kinetics is modelled here by replacing the Butler-Volmer prefactor, $j_0$ with a sigmoid function of the form
\begin{equation}
  j_0(\xi_N) = \frac{j_0}{1+\exp((\xi_N - \xi_\text{max})/l)}, 
  \label{eq:BV1}
\end{equation}
See Fig \ref{fig:sigmoid}. This represents one aspect coupling mechanics with electrochemical degradation. 

\subsection{Stress dependent kinetics}
In addition to electrochemical degradation due to interface fracture, we account for stress dependent kinetics following Ref \cite{Garikipati2001Rao}. The diffusivity in the solid electrolyte can be stress dependent. Tensile stresses cause a local expansion of the solid electrolyte's crystal structure. This typically lowers the energy barrier for diffusive hops of Li or Li$^+$ and vacancies and enhances diffusion. In terms of transition state theory, the activation energy for diffusion is modified by a work-like term of the form $\boldsymbol{\sigma}\colon\boldsymbol{V}_\text{D}$ \cite{garikipati2006continuum,puchala2008elastic,ganser2019extended}, where $\boldsymbol{\sigma}$ is the Cauchy stress and $\boldsymbol{V}_\text{D}$ is the activation volume tensor for diffusion. In the absence of detailed experimental or first principles computations on its tensorial character, we adopt an isotropic model: $\boldsymbol{V}_\text{D} = V_\text{D}\boldsymbol{1}$. This leads to the following stress-dependent diffusivity \cite{garikipati2006continuum,ganser2019extended}, which is applied to $D_\text{Li}$ and $D_{\text{Li}^+}$ in \eref{eq:Liflux} and \eref{eq:Li+flux}, respectively:
\begin{equation}
  D(\boldsymbol{\sigma})=D_0 \exp \left( \frac{\text{tr}[\boldsymbol{\sigma}]V_\text{D}}{kT} \right).
  \label{eq:actvoldiff}
\end{equation}
The reaction rate also has a stress dependence with similar origins. However, since charge transfer occurs on the interface, $\Gamma$, we model the corresponding activation volume tensor to have the form $\boldsymbol{V}_\text{R} = V_\text{R}\boldsymbol{n}\otimes\boldsymbol{n}$, which differs from the hydrostatic stress-dependence in Ref \cite{ganser2019extended}. The prefactor in the Butler-Volmer model is further modified to
\begin{equation}
  j_0(\boldsymbol{\sigma})=j_{0} \exp \left( \frac{\boldsymbol{n}\cdot\boldsymbol{\sigma}\boldsymbol{n}V_\text{R}}{kT} \right).
  \label{eq:BV2}
\end{equation}

The final form of the charge transfer kinetics with fracture-induced degradation and stress-dependence is obtained from \eref{eq:BV0}, \eref{eq:BV1} and \eref{eq:BV2}:

\begin{align}
    \Bj_{\text{Li}^+}\cdot \Bn^{+} + \Bj_\text{Li} \cdot \Bn^{-} =& \frac{j_0}{1+\exp((\xi_N - \xi_\text{max})/l)} \nonumber\\
    &\times\exp \left( \frac{\boldsymbol{n}\cdot\boldsymbol{\sigma}\boldsymbol{n}V_\text{R}}{kT} \right)\nonumber\\
    &\times\left(\text{exp}\left(\frac{\alpha_aF}{R\theta}(\phi_\text{p}^- -\phi_\text{e}^+ -U)\right)-\text{exp}\left(-\frac{\alpha_aF}{R\theta}(\phi_\text{p}^- -\phi_\text{e}^+ -U)\right)\right)\quad\text{on }\Gamma.
    \label{eq:BVfull}
\end{align}

\section{Efficient generation of multi-particle configurations}\label{sec:meshgen}

\begin{figure}[h!]
        \centering
        \includegraphics[width=0.8\textwidth]{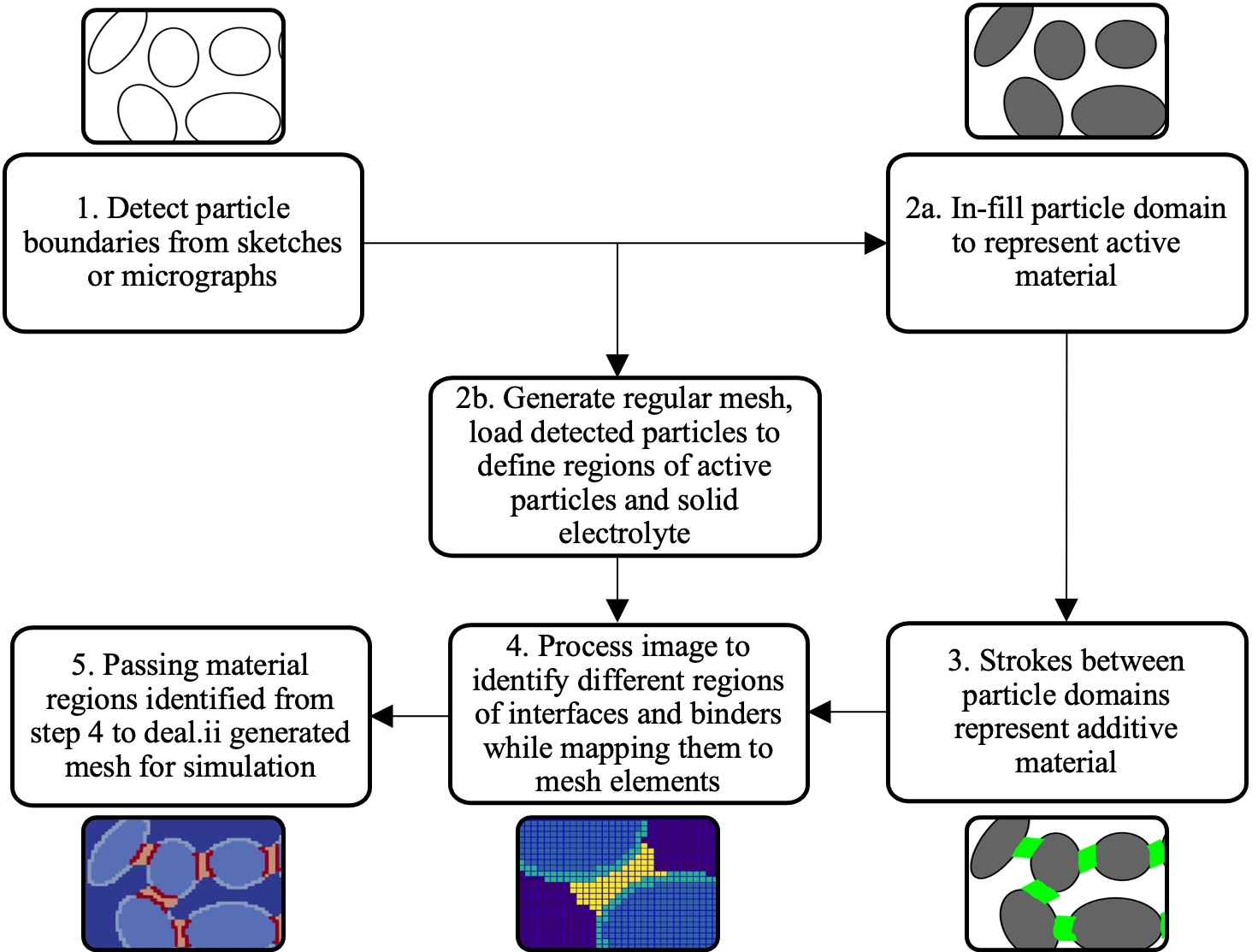}
        \caption{The image-based mesh generation workflow.}
        \label{fig:imagebasedmeshsetup}
      \end{figure}
\begin{figure}[h!]
    \centering
    \includegraphics[width=0.9\linewidth]{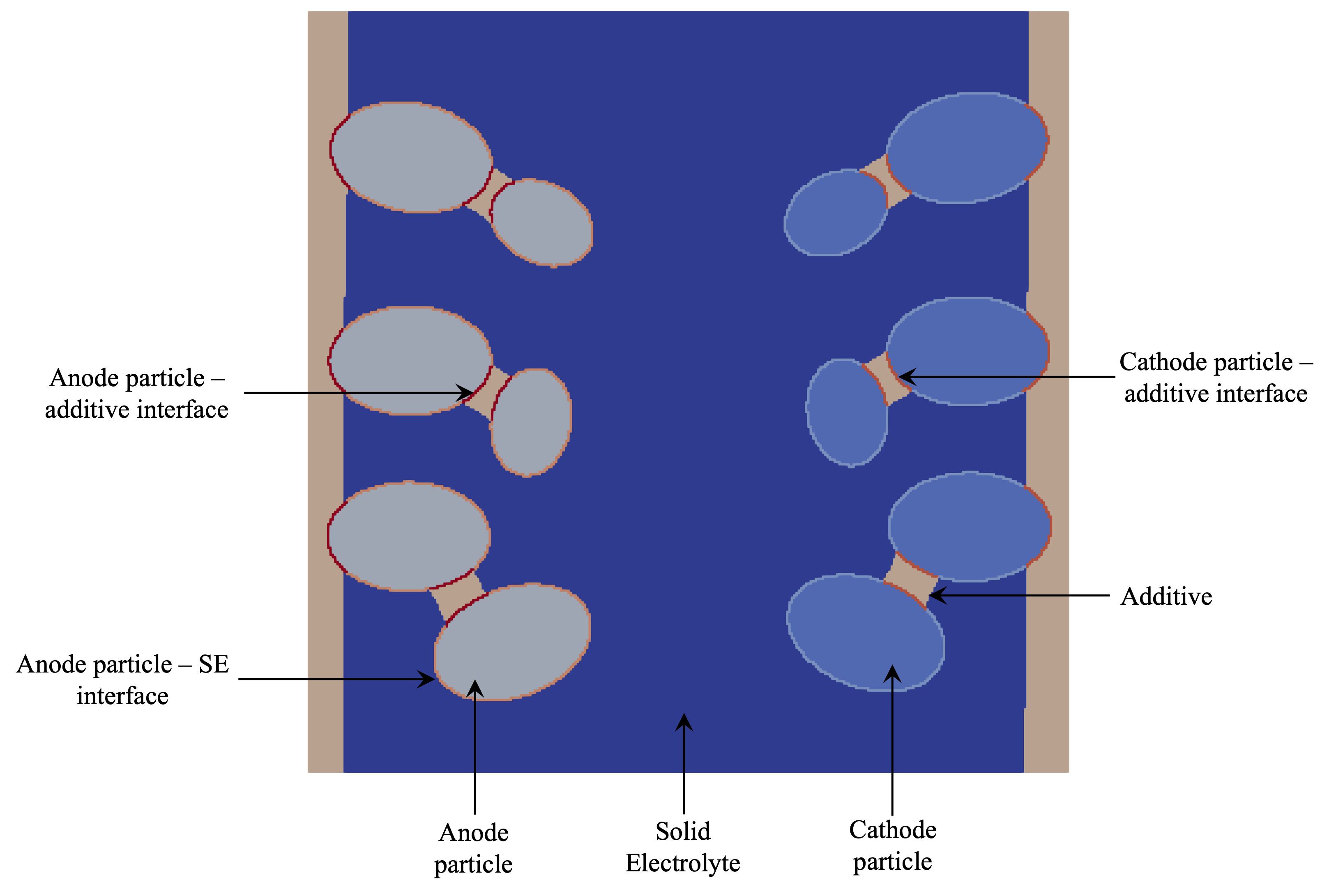}
    \caption{Material labels of the multi-particle configuration generated in Fig \ref{fig:imagebasedmeshsetup}.}
    \label{fig:material_id}
\end{figure}

Fig \ref{fig:imagebasedmeshsetup} illustrates the workflow by which we generate multi-particle configurations. While this example begins with the definition of elliptical particles in an idealized geometry, it can be extended to work with micrographs of active particles in a solid state electrolyte by an easy replacement of the steps in the bottom row of Fig \ref{fig:imagebasedmeshsetup}.  The image processing feature embedded in the workflow allows the seamless recognition of arbitrary particle shapes and fits an elliptical interface around each particle. The workflow also recognizes the binder strokes between the particles represented by the Chartreuse green color in the top row of Fig \ref{fig:imagebasedmeshsetup} and defines the sub-domain for additive material. Given Cartesian mesh data which includes only the number of elements in each direction, the workflow then maps each element to sub-domains recognized from the image. Figure \ref{fig:material_id} illustrates these six sub-domains, namely (i) Anode particle (ii) Cathode particle (iii) Solid Electrolyte (SE) (iv) Anode-SE interface  (v) Cathode-SE interface, and (vi) Additive. As discussed in \ref{sec:embeddedintfc}, the mathematical treatment of the interfaces allows the meshes around these sub-domains to be Cartesian with the quadrilateral elements holding information for each sub-domain. 

\section{Multiphysics computations on solid state batteries}\label{sec:simulation}
We demonstrate the multiphysics computational framework on a solid state battery chemistry, first by considering idealized configurations with single anode and cathode particles, and next on a multi-particle configuration. With the single anode/cathode particle cases, we separate out the effects of stress-dependent kinetics and interface fracture on Li and Li$^+$ concentration fields, electrode and electrolyte potentials. We consider a single discharge-charge cycle, postponing a detailed study of cycling to a future communication.

We adopt the St. Venant-Kirchhoff model of nonlinear elasticity. The first Piola-Kirchhoff stress, expressed in terms of the elastic deformation gradient and Green-Lagrange strain, $\bar{\boldsymbol{E}}^\text{e}$, is:
\begin{equation}
    \boldsymbol{P} = \bar{\boldsymbol{F}}^\text{e}\left(\lambda\text{tr}[\bar{\boldsymbol{E}}^\text{e}] \boldsymbol{1} + 2G\bar{\boldsymbol{E}}^\text{e}\right)\bar{\boldsymbol{F}}^{\text{c}^{-1}}
    \label{eq:SVK-PK}
\end{equation}
for Lam\'{e} parameter, $\lambda$ and shear modulus $G$ for the respective materials (electrolyte, active particles, binder) obtained from the reported Young's modulus $E$ and Poisson ratio $\nu$ for the solid materials: $\lambda = \nu E/((1+\nu)(1-2\nu))$ and $G = E/(2(1+\nu))$.

Equation \eref{eq:fc} is now modified so that the chemical component of the regular part of the deformation gradient $\bar{\boldsymbol{F}^\text{c}}$ is specified by the intercalation function, $g(c_\text{Li})$, which is defined as: 
\begin{equation}
    g(c_\text{Li}) = \left(\left(\frac{c_\text{Li}-c_0}{c_\text{Li}^\text{max} - c_\text{Li}^\text{min}}\right)r_\text{s} + 1\right)^{3}
    \label{eq:gc}
\end{equation}

where $c_0$ is the initial Li concentration and the swell ratio $r_\text{s}$ is obtained from the maximum volume change $\Delta V$ reported for the individual electrodes: $r_\text{s} = \left(1 + \Delta V)\right)^{1/3} -1$. 

\begin{table}
\centering
\caption{Electro-chemo-mechanical parameters.}
\label{table:paramstable}
\setlength{\tabcolsep}{4.5pt} 
\renewcommand{\arraystretch}{1.5} 
\begin{tabular}{|c c c c c c|}
\hline

\textbf{Symbol} & \textbf{Name} & \textbf{Unit} & \textbf{Anode} & \textbf{SE} & \textbf{Cathode} \\ [0.5ex] 
\hline
\multicolumn{6}{|c|}{\textbf{Constant}}\\ 

$F$ & Faraday’s constant & pC/pmol & - & 96487 & -\\ 
$R$ & Universal gas constant & pJ/(pmol·K) & - & 8.3143 & -\\ 
$\theta$ & Temperature  & K & - & 298 & -\\ 

\multicolumn{6}{|c|}{\textbf{Cell Geometry}}\\ 
$L$ & Cell length &  $\mu\text{m}$  & - & 80 & -\\ 
$W$ & Cell width & $\mu\text{m}$ & - & 80 & -\\ 
 
\multicolumn{6}{|c|}{\textbf{Electrochemical Parameters}}\\ 
$\alpha_a$ & Transfer coeff \cite{Wang+Garikipati2017}\cite{Wang+Garikipati2018}& - & 0.5 & - & 0.5\\ 

$\kappa_\text{e}$ & Conductivity of $\text{Li}^+$ \cite{hayashi2001preparation} & p$\left(\Omega\mu\text{m}\right)^{-1}$ & - & $1.6\times10^{4}$ & -\\ 
$D_\text{Li}$ & Diffusivity of Li \cite{Wang+Garikipati2017}\cite{Wang+Garikipati2018}& $\mu\text{m}^2$/s & 0.5 & - & 0.5\\ 

$D_{\text{Li}^+}$ & Diffusivity of $\text{Li}^+$ \cite{de2018analysis}  & $\mu\text{m}^2$/s & - & 1000 & -\\ 
$t^+$ & Transference number\cite{Wang+Garikipati2017}\cite{Wang+Garikipati2018} & - & - & 0.2 & -\\ 
$c_\text{Li}^\text{max}$ & Maximum Li conc (est.) \cite{Wang+Garikipati2017}\cite{Wang+Garikipati2018} &  pmol/$\mu\text{m}^3$ & 0.0262605 & - & 0.03675\\ 
$c_\text{Li}^\text{min}$ & Minimum Li conc (est.) \cite{Wang+Garikipati2017}\cite{Wang+Garikipati2018}&  pmol/$\mu\text{m}^3$ & 0.000574 & - & 0.000825\\ 
$c_\text{Li}^\text{0}$ & Initial Li conc &  pmol/$\mu\text{m}^3$ & 0.0262605 & - & 0.000825\\ 
$c_{\text{Li}^+}^0 $ & Initial  $\text{Li}^+$ conc (est.)  &  pmol/$\mu\text{m}^3$  & - & 0.002 & -\\ 
$\text{V}_\text{R}, \text{V}_\text{D}$ & Activation volumes\cite{XingZhang2020stress}  &  $\text{m}^3$  &  $5.807\times10^{-30}$ & -  & $5.807\times10^{-30}$\\ 
$\Delta V$ & Maximum volume change \cite{takami2011lithium}\cite{ghosh2020role}\cite{iniguez2022structure} \cite{luo2016situ}& \% & 0 & - & 1.9 \\ 

\multicolumn{6}{|c|}{\textbf{Elasticity Parameters}}\\ 
$E$ & Young’s modulus \cite{fergus2010ceramic} \cite{baranowski2016multi} \cite{cheng2017elastic}  & GPa & 30 & 10 & 190\\ 
$\nu$ & Poisson’s ratio & - & 0.3 & 0.3 & 0.3\\ 
$f_t$ & Fracture strength \cite{okamura1995ceramic} \cite{gupta1994recent} & GPa & 0.3 & - & 0.3\\ 
\hline

\end{tabular}
\end{table}

\subsection{An idealized single anode/cathode particle configuration}\label{sec:singleparticle}

We consider a domain $80\times 80\;\mu$m to define the cell with single anode and cathode particles. In all the figures that follow the anode particle is to the left and the cathode to the right. The simulations are of a Lithium titanate (LTO) anode, LCO cathode, and $\beta\text{-Li}_3\text{PS}_4$ solid electrolyte. The properties of these materials and parameters used for the simulations have been summarised in Table \ref{table:paramstable}. To demonstrate the effect of stress-dependence on the kinetics in the absence of fracture, we present the examples of (i) stress-independent kinetics (ii) stress-dependent diffusion (iii) stress-dependent reaction, and (iv) Combined effects of stress-dependent diffusion, and reaction. We also discuss the effect of stress and stress-induced fracture on the charge transfer process. 

\begin{figure}[h!]
        \centering
        \includegraphics[width=0.8\textwidth]{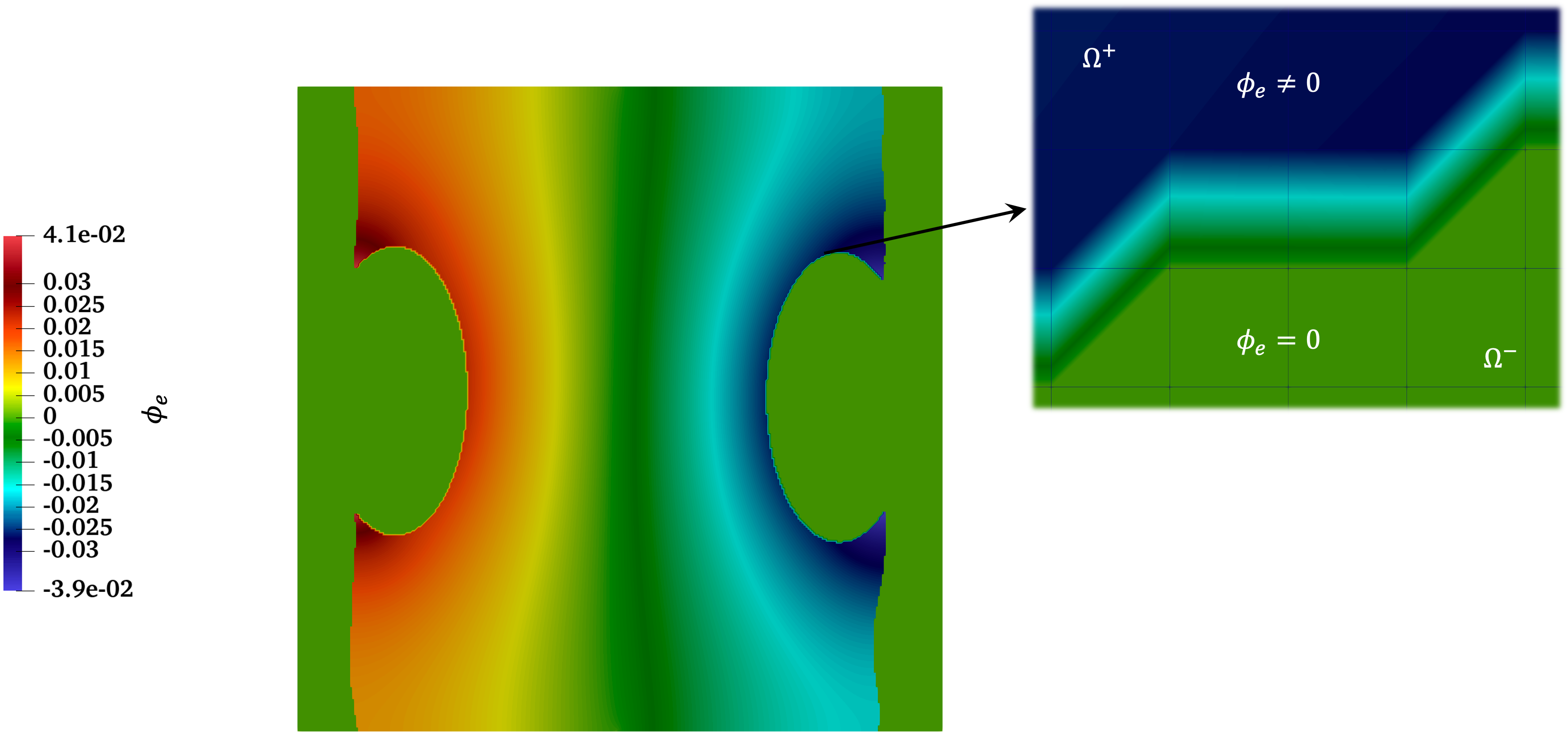}
        \caption{The discontinuous electric potential field $\phi_\text{e}$ (V) at the end of the first discharge from a computation run with stress-independent kinetics and fracture suppressed.}
        \label{fig:discontinuousphi_e}
\end{figure}

\begin{figure}[h!]
        \centering
        \includegraphics[width=0.8\textwidth]{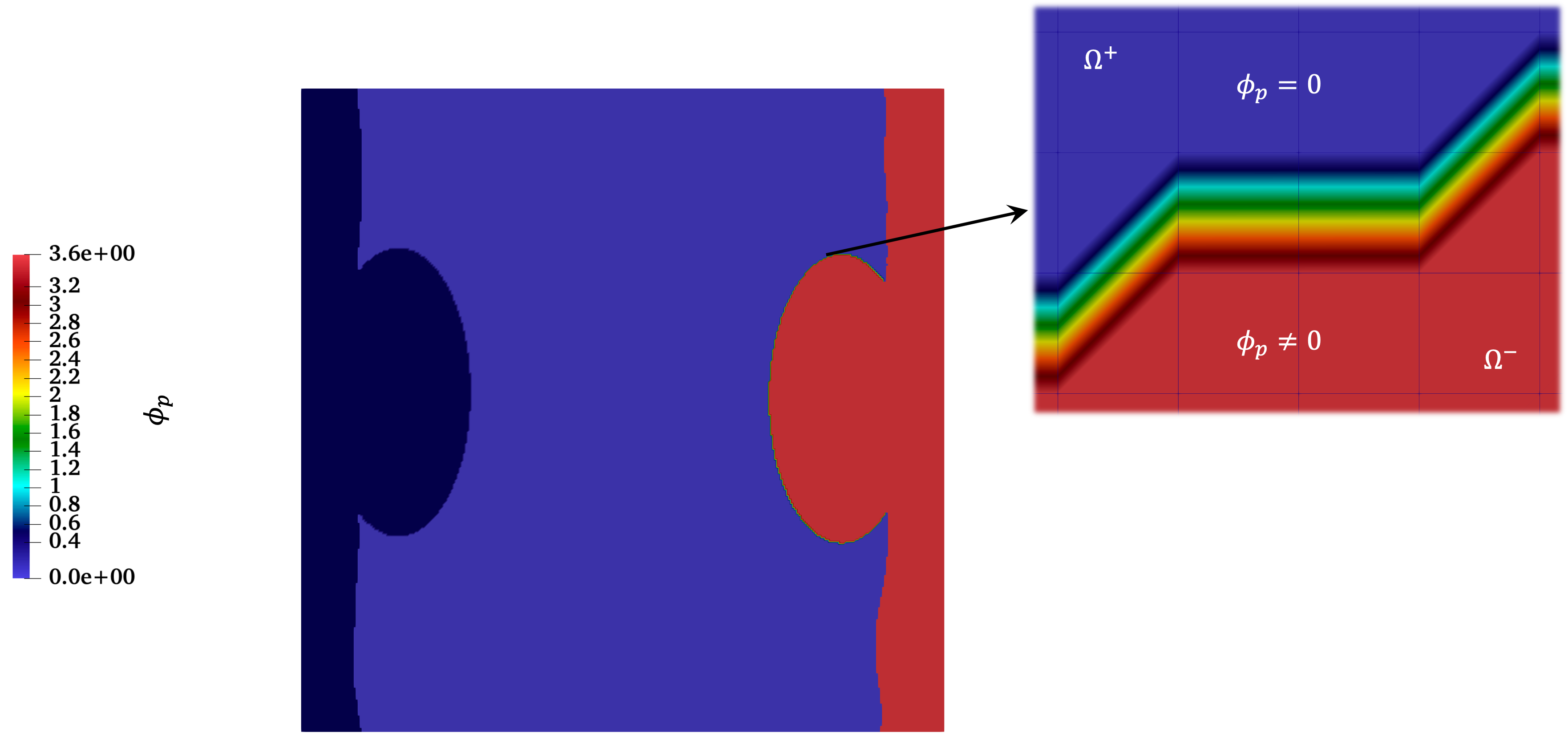}
        \caption{The discontinuous electric potential field $\phi_\text{p}$ (V) at the end of the first discharge from a computation run with stress-independent kinetics and fracture suppressed.}
        \label{fig:discontinuousphi_p}
\end{figure}

\begin{figure}[h!]
        \centering
        \includegraphics[width=0.08\textwidth]{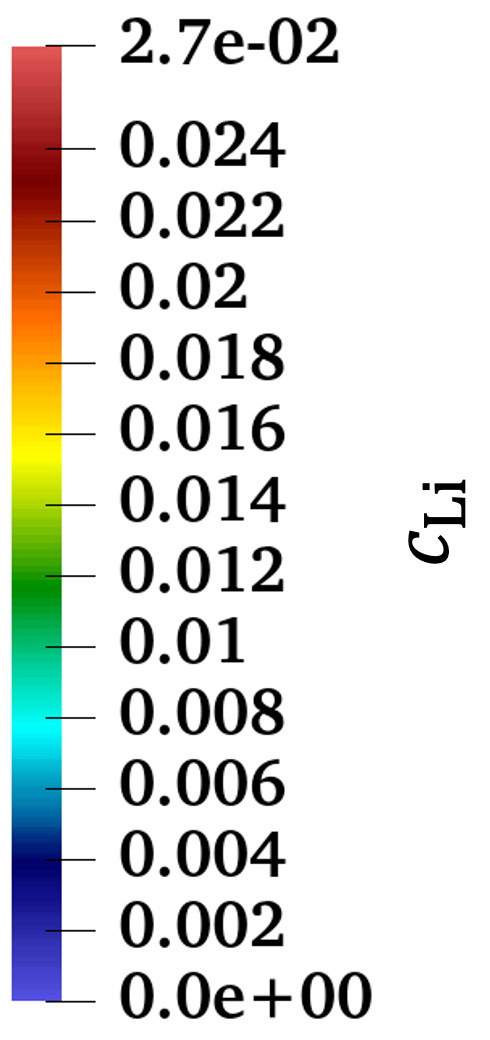}
        \includegraphics[width=0.3\textwidth, trim=150mm 50mm 150mm 20mm]{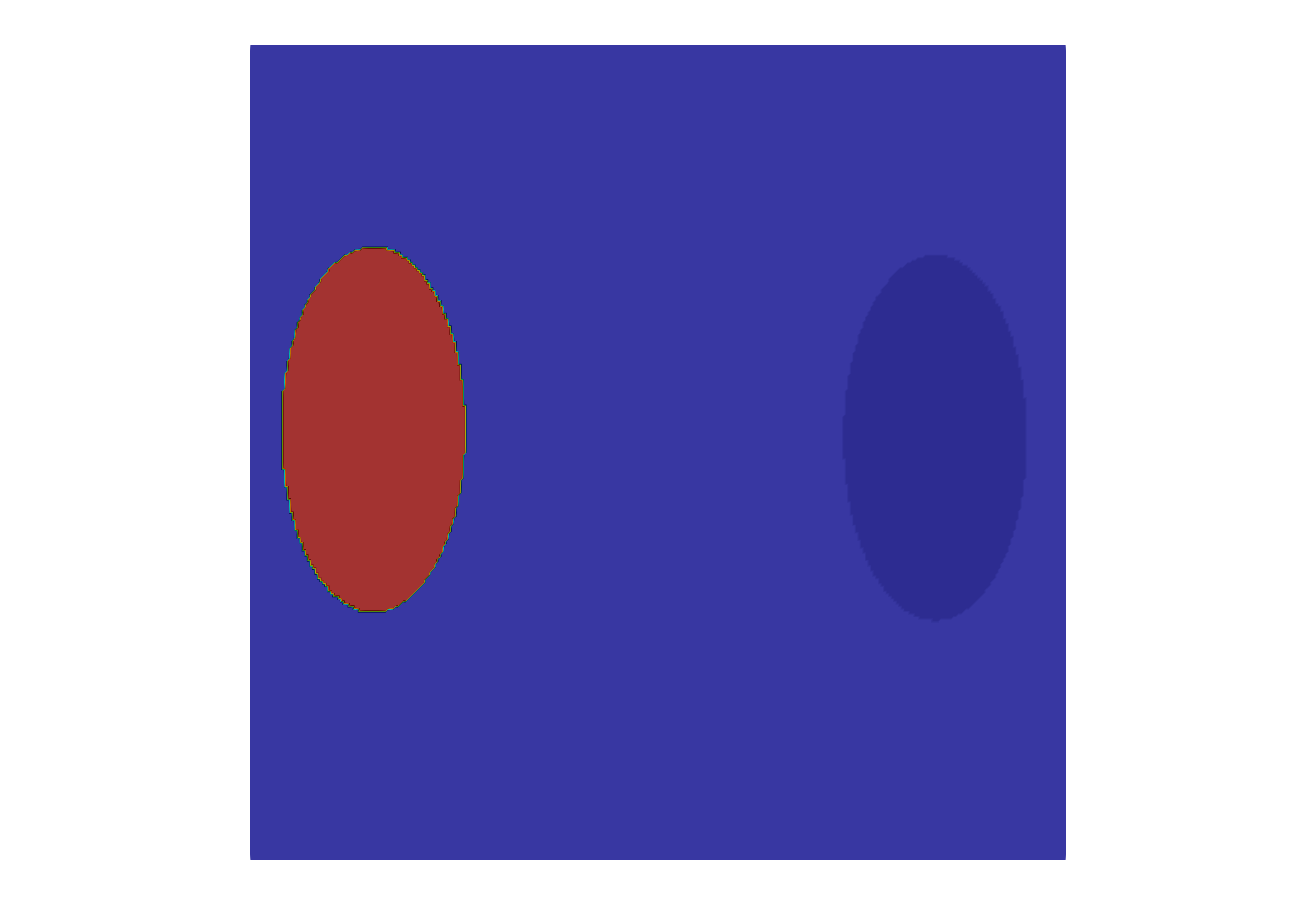}
        \includegraphics[width=0.3\textwidth, trim=150mm 50mm 150mm 20mm]{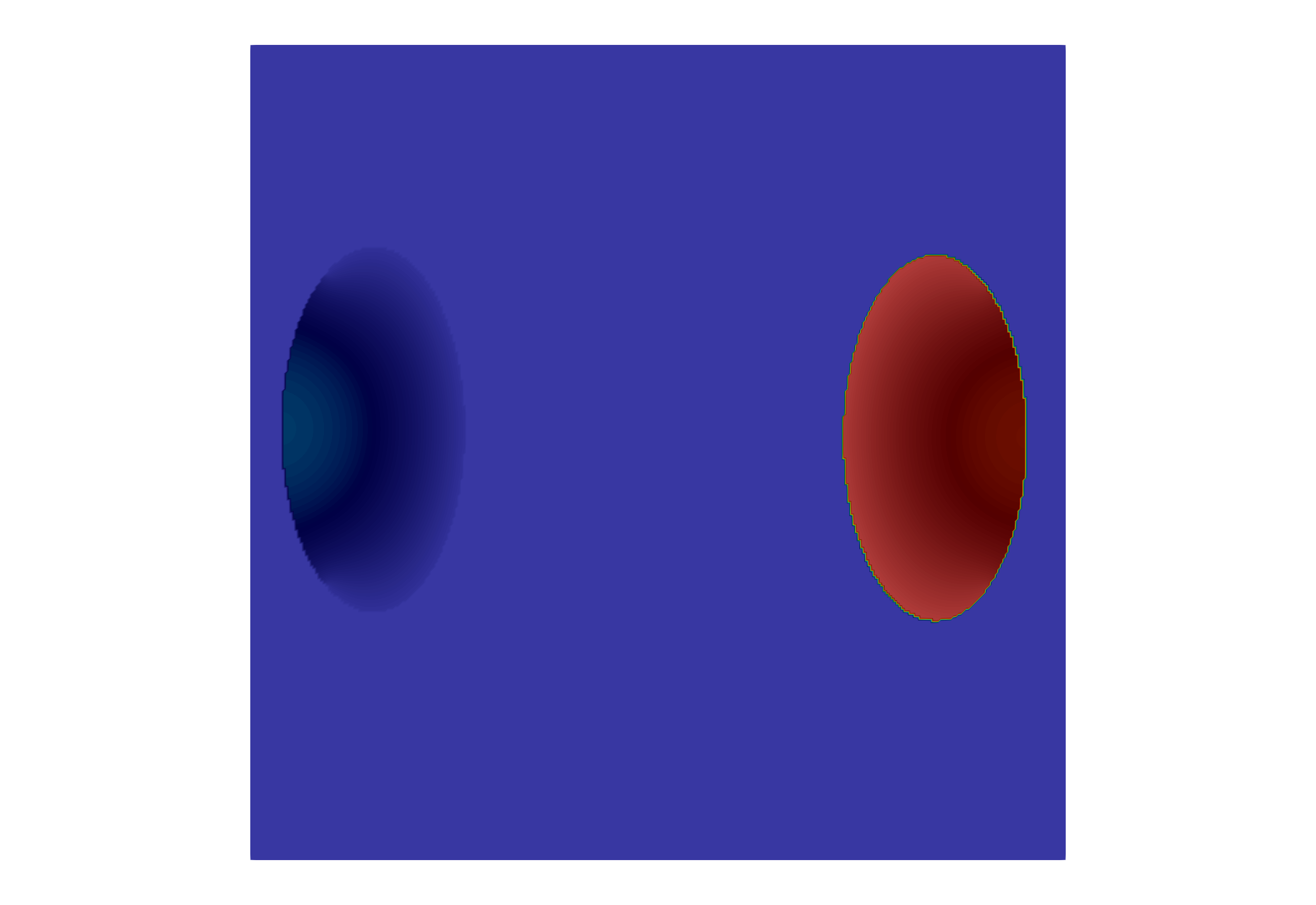}
        \includegraphics[width=0.3\textwidth, trim=150mm 50mm 150mm 20mm]{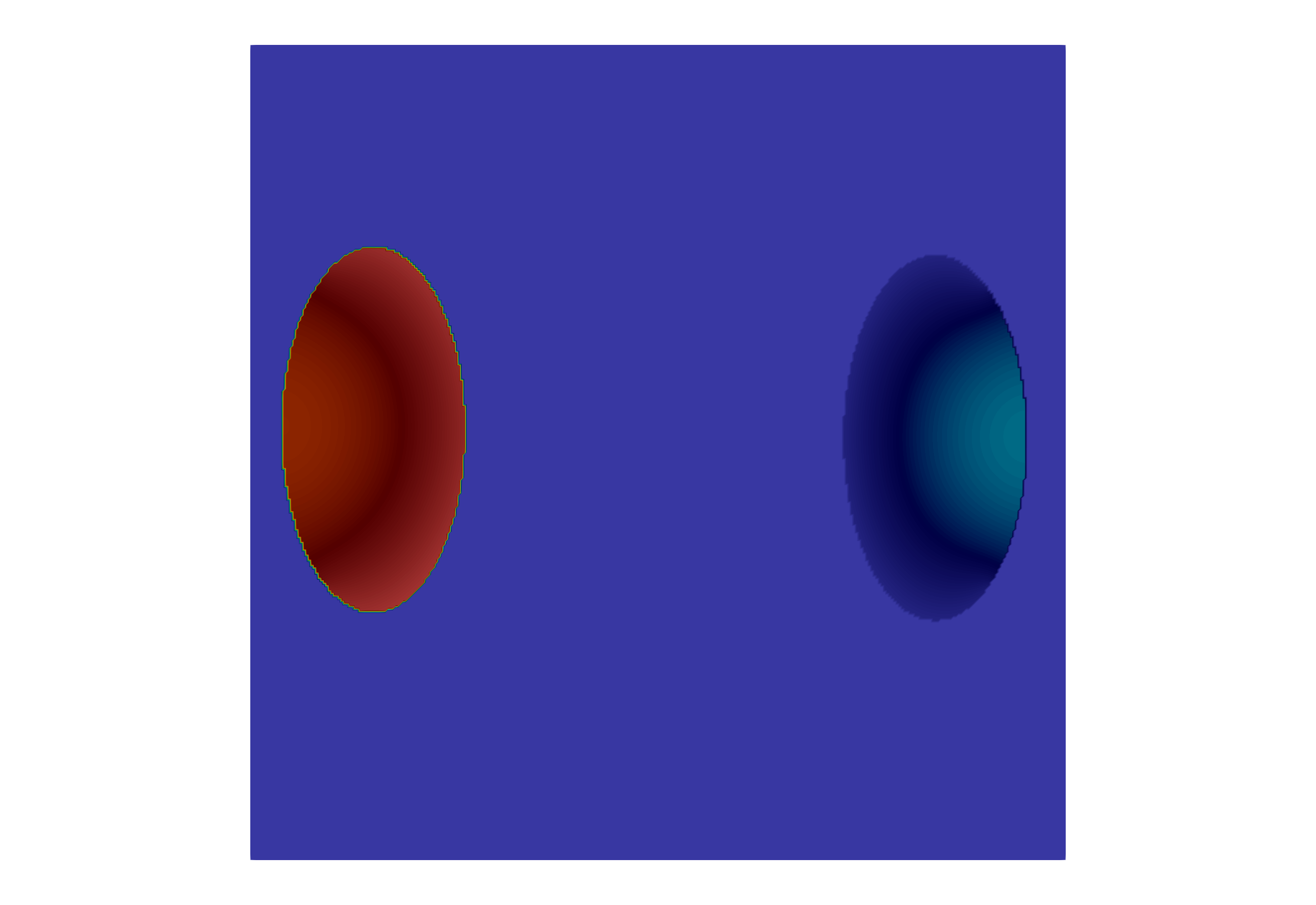}
        \caption{Li concentration field ($\text{pmol/}\mu\text{m}^3$) in a single-particle cell with stress-independent kinetics and no fracture. Left: initial state. Middle: end of $1^\text{st}$ discharge. Right: end of $1^\text{st}$ charge}
        \label{fig:singleparticlenostressnofracture}
      \end{figure}

Fig \ref{fig:singleparticlenostressnofracture} shows the distribution of Li initially, at the end of the discharge and at the fully recharged state, computed with a version of the model in which stress effects on kinetics have been suppressed in Eqs (\ref{eq:actvoldiff}--\ref{eq:BVfull}) by setting the  activation volumes $V_\text{D}, V_\text{R} = 0$. Figure \ref{fig:discontinuousphi_e} shows the jump in electrolyte potential $\phi_\text{e}$ at $\Gamma$ from non-zero values in the electrolyte to zero in the active particles. The discontinuity is imposed by the basis function $M_\Gamma$. However, in the plot, this discontinuity undergoes a smooth interpolation over a single element $\Omega_e^\Gamma$. Similarly, the electrode potential $\phi_\text{p}$ also suffers a jump from non-zero in the active particles to zero everywhere in the electrolyte, as shown in Figure \ref{fig:discontinuousphi_p}.

\begin{figure}[h!]
        \centering
        \includegraphics[width=0.8\textwidth]{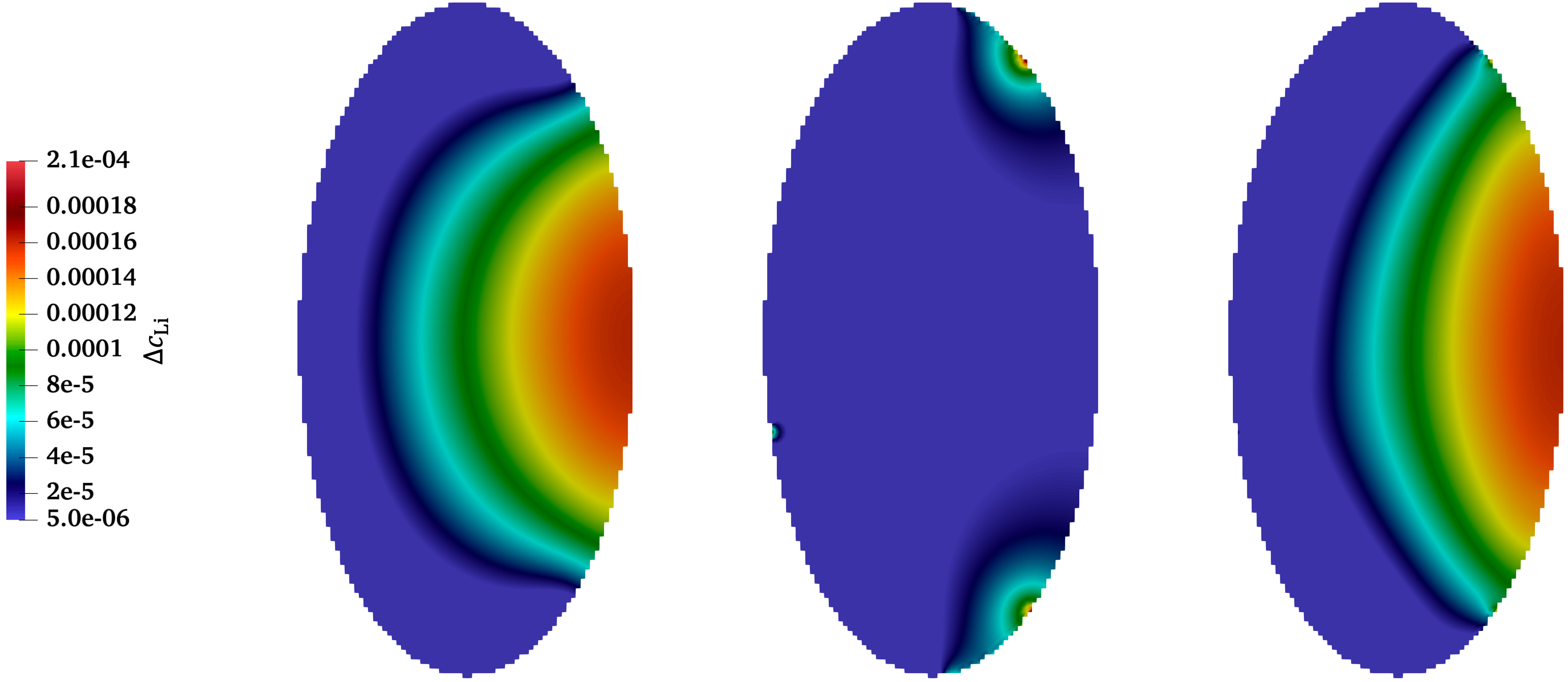}
        \caption{Comparison plots showing the distribution of the differences $\Delta c_{\text{Li}}$ ($\text{pmol/}\mu\text{m}^3$) in the cathode particle compared to the baseline simulation with stress-independence in the kinetics. Left: stress-dependent diffusion. Middle: stress-dependent reaction. Right: combined effect of stress-dependent diffusion and reaction. All results are at the end of the first discharge and in the absence of interfacial fracture.}
        \label{fig:singleparticlestressvsnostress}
\end{figure}

To represent the stress effects on the distribution of Li, the simulation results shown in Fig \ref{fig:singleparticlenostressnofracture} are considered as a baseline to which stress effects on diffusion and/or reaction rate are compared in the absence of fracture. We draw attention to the discontinuous $c_\text{Li}$ field. It has the same smoothed interpolation of discontinuities seen in Figures \ref{fig:discontinuousphi_e} and \ref{fig:discontinuousphi_p}. For comparison of stress effects on the kinetics, we define $\Delta c_{\text{Li}}$ as the deviation in Li concentration, $c_{\text{Li}}$, from the baseline case with stress-independent kinetics. Fig \ref{fig:singleparticlestressvsnostress} shows a detailed comparison of the distribution of $\Delta c_{\text{Li}}$ in the cathode active particle at the end of the discharge half-cycle. The distribution shown in the left particle of Fig \ref{fig:singleparticlestressvsnostress} is obtained by setting $V_\text{D} = 5.807\times10^{-30} \text{ m}^3$\cite{XingZhang2020stress}, $V_\text{R} = 0$, activating stress-dependent diffusion and suppressing the stress-dependent reaction rate. Due to the Dirichlet boundary condition near the current collector and intercalation strain in stiff cathode particles, the tensile stresses are higher where the current collector connects to the solid electrolyte or an active particle. Eq \ref{eq:actvoldiff} indicates that for $V_\text{D}> 0$, a state of tensile hydrostatic stress  will enhance transport by diffusion. As a result, both Li and Li$^+$ transport by diffusion is enhanced in this region providing more Li in the core of the cathode particle as seen in the left particle of Fig \ref{fig:singleparticlestressvsnostress}. Furthermore, the distribution of $\Delta c_{\text{Li}}$ in the middle particle of Fig \ref{fig:singleparticlestressvsnostress} can also be explained as a stress-driven enhancement. In this case, we activate stress-dependent reaction and suppress any stress effects on diffusion by setting $V_\text{R} = 5.807\times10^{-30} \text{ m}^3$\cite{XingZhang2020stress}, $V_\text{D} = 0$. For $V_\text{R} > 0$, Eq \ref{eq:BV2} indicates that the higher tensile stresses in the interface near the current collector accelerate the interface charge transfer kinetics which can be seen in the form of two hotspots of additional Li in the middle particle of Fig \ref{fig:singleparticlestressvsnostress}. Lastly, we activate the stress effects on both diffusion and reaction by setting $V_\text{R} =  V_\text{D} = 5.807\times10^{-30} \text{ m}^3$. As expected, the distribution of $\Delta c_{\text{Li}}$ shown in the right particle of Fig \ref{fig:singleparticlestressvsnostress} is the result of enhanced diffusion and accelerated interface charge transfer kinetics and manifests as additional Li on the interface near the current collector in comparison with the previous two cases.

\begin{figure}[h!]
        \centering
        \includegraphics[width=0.12\textwidth]{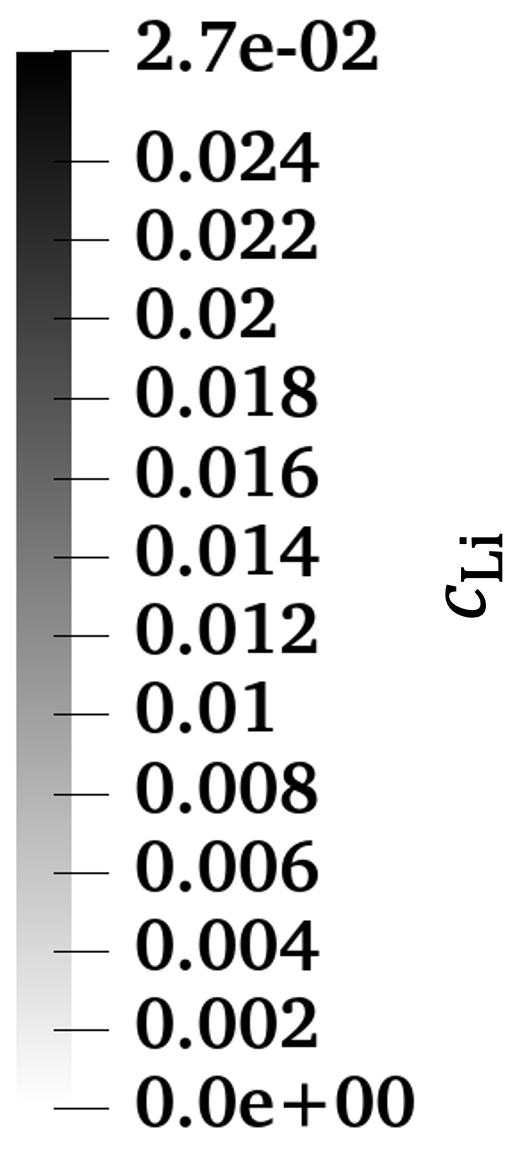}
        \hspace{1 cm}
        \includegraphics[width=0.8\textwidth]{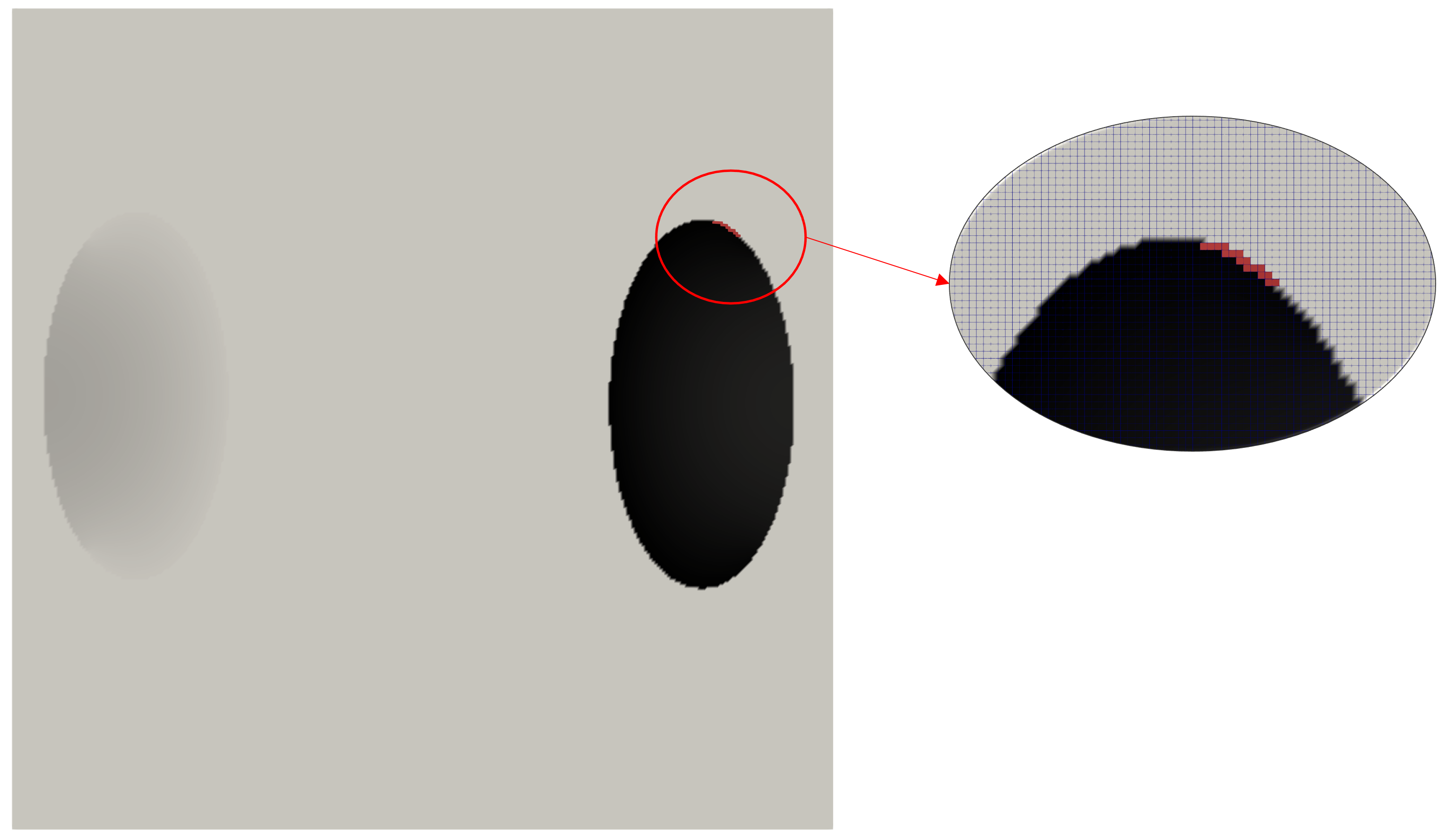}
        \caption{Fractured elements in red on the interface between solid electrolyte and cathode active particles at the end of the first discharge.}
        \label{fig:singleparticlefractureelems}
\end{figure}

\begin{figure}[h!]
        \centering
        \includegraphics[width=1\textwidth]{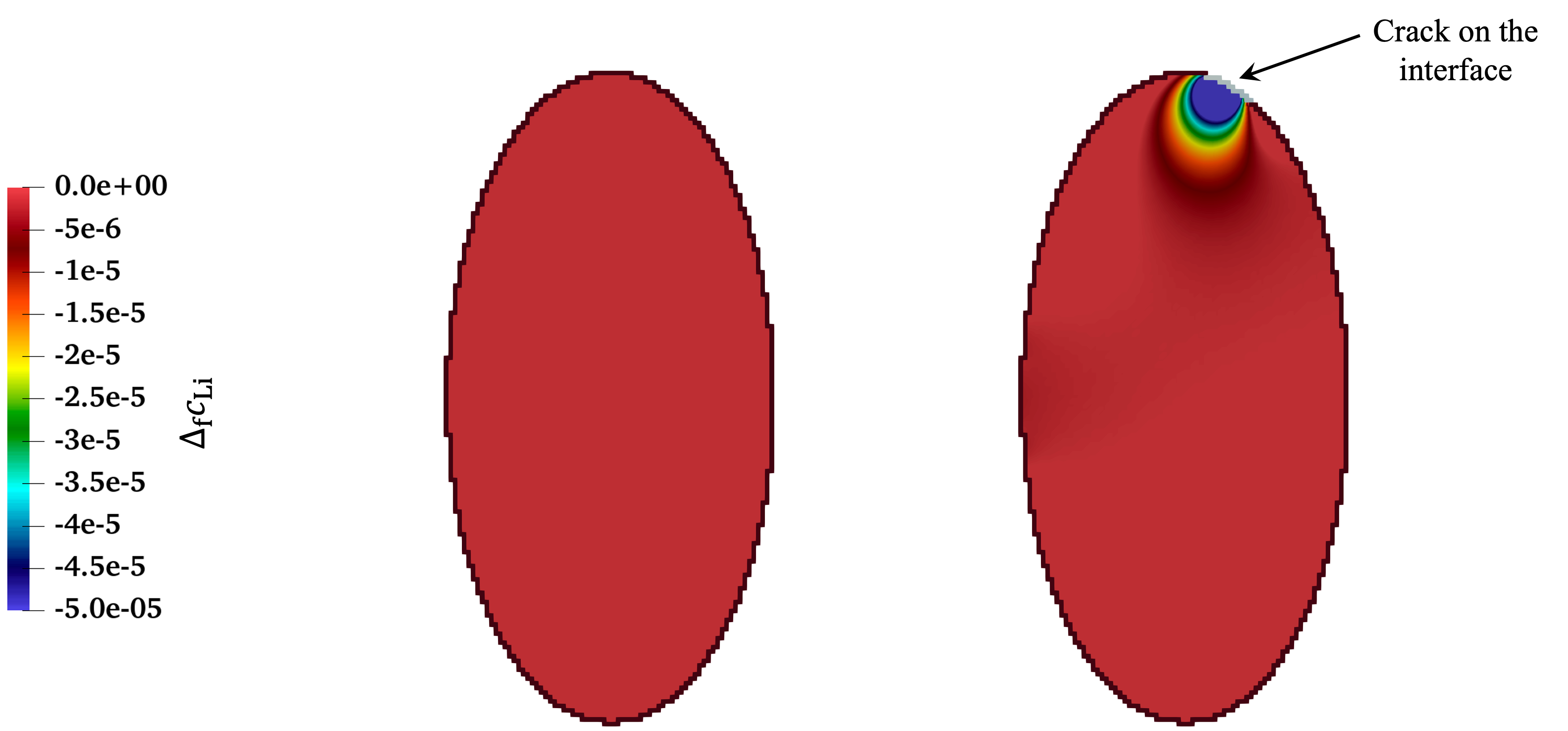}
        \caption{Comparison of deviation in Li concentration  $\Delta_\text{f} c_{\text{Li}}$ ($\text{pmol/}\mu\text{m}^3$) in the cathode active particle between the fracture and fracture-suppressed cases at the end of the first discharge. Left:  fracture-suppressed. Right: Fracture with cracked elements in white.}
        \label{fig:singleparticlefracturevsnofracture}
\end{figure}

The central electrochemical phenomenon that we seek to capture with the fracture model is the degradation of charge transfer. To demonstrate this effect we extend the computations by activating the fracture model. 
Figure \ref{fig:singleparticlefractureelems} shows in red the  elements that fractured  during the discharge cycle. As discussed above, the higher tensile stresses on the interface near the current collector result in crack initiation at the top edge of the elliptical particle and propagation along the interface nearer to the current collector. To show the degradation in Li transfer across the fractured interface as a result of Eq \ref{eq:BV1}, Fig \ref{fig:singleparticlefracturevsnofracture} presents the distribution of differences $\Delta_\text{f} c_{\text{Li}}$ in the cathode particle, which is the deviation of $c_{\text{Li}}$ from the case with stress-dependent-kinetics but fracture suppressed. Notably, even though the cracking is along a relatively small contour length of the interface, the opening induces degradation of the charge transfer kinetics as shown in Fig \ref{fig:singleparticlefracturevsnofracture}. Since the traction decreases with crack opening we expect that the tensile stress levels are lower in the particles and electrolyte. With additional discharge-charge cycles (not simulated here) this could lead to lower diffusivity enhancement, fewer Li$^+$ ions arriving at the interface, and a further decrease in Li levels within the cathode particle. 

In addition to the loss in Li transfer across the interface in Fig \ref{fig:singleparticlefracturevsnofracture}, we also demonstrate the increase in the effective internal resistance $R_{\text{eff}}$ due to the interface opening. To evaluate $R_{\text{eff}}$, we consider the drop in potential difference $\Delta\phi = U_{\text{ocv}} - V_T$. Here, the terminal voltage $V_\text{T}$ is the potential difference across the boundaries of the cell given by $V_\text{T} = \phi_{\text{p}_0} - \phi_{\text{p}_L}$, where the subscripts $0,L$ correspond to the left and right terminals of the cell. The open circuit voltage is $U_{\text{ocv}} = U^+ - U^-$. The half-cell potentials for the cathode and anode, respectively, $U^+ \text{and } U^-$ respectively are written as fits\cite{Wang+Garikipati2018}::
\begin{subequations}
\begin{align}
U^+ &= \frac{-0.0923 - 7.8680x + 50.0722x^2 - 122.2816x^3 + 82.9851x^4 + 140.2939x^5 - 374.7350x^6 + 403.2464x^7 - 221.1915x^8 + 49.3393x^9}{-0.0218 - 1.9007x + 11.7264x^2 - 28.7848x^3 + 27.5427x^4 - 8.6343x^5}\label{eq:uplus} \\
U^- &= 0.2657 + 0.5551 e^{-178.9799x} - 0.0124 \tanh\left(\frac{x - 0.5573}{0.0282}\right) - 0.0117 \tanh\left(\frac{x - 0.2393}{0.0486}\right) - 0.0129 \tanh\left(\frac{x - 0.1749}{0.0348}\right) - 0.0501 \tanh\left(\frac{x - 0.99}{0.0245}\right) - 0.0353x - 0.0119 \tanh\left(\frac{x - 0.1299}{0.0198}\right) - 0.1526 \tanh\left(\frac{x - 0.03}{0.0227}\right) \label{eq:uminus}
\end{align}
\end{subequations}
where $x = \bar{c}_\text{Li}/c_\text{Li}^\text{max}$, the ratio of the volume averaged concentration to the maximum concentration for the respective electrode.  The current density is the applied charge flux in the single particle simulations, $i_\text{ext} = 15\;\text{pA}\mu\text{m}^{-2}$. As relevant cell dimensions, for instance in a ``jelly-roll" structure, we use the cell's charge transfer area $A_\text{cell} = 10^3\;\text{cm}^2$, leading to the applied current, $i_\text{app} = 1.5\;A$.  Ohm's law for an effective resistance $R_\text{eff}$ is $i_\text{app}R_\text{eff} = \Delta\phi$. Using the values of $x$, as defined above, obtained from the computations at the end of discharge for the cathode with/without fracture and the corresponding anode conditions, we find, $U_\text{OCV}^\text{nofrac} = 3.4110\;\text{V}$, $V_\text{T}^\text{nofrac} = 3.0515\;\text{V}$, yielding $\Delta\phi^\text{nofrac} = 0.3595\;\text{V}$ and $R_\text{eff}^\text{nofrac} = 0.2396\;\Omega$. For the fractured single particle, these quantities as extracted from the computations were: $U_\text{OCV}^\text{frac} = 3.4110\;\text{V}$, $V_\text{T}^\text{frac} = 3.0484\;\text{V}$, yielding $\Delta\phi^\text{frac} = 0.3626\;\text{V}$ and $R_\text{eff}^\text{frac} = 0.2418\;\Omega$. Notably, while the computations clearly demonstrate a degradation of charge transfer into the cathode at the end of discharge (Figure \ref{fig:singleparticlefracturevsnofracture}), the average concentration difference is small, and the $U_\text{OCV}$ is unchanged to the fourth decimal. However, the resulting potential field $\phi_\text{p}$, being coupled with $c_\text{Li}$ via the field equations (\ref{eq:weakmasstransp}-\ref{eq:weakelectroelectrolyte}) and (\ref{eq:BVfull}) is disturbed sufficiently due to fracture leading to a difference in $\Delta\phi$ and therefore in $R_\text{eff}$. While small, we note that fracture at the cathode-electrolyte interface does lead to an increased Ohmic resistance by $2.2\;\text{m}\Omega$. This will grow as fracture progresses with cycling.

\subsection{Simulations with multiple particles}\label{sec:multipleparticle}

\begin{figure}[h!]
    \centering
    \includegraphics[width=1\linewidth]{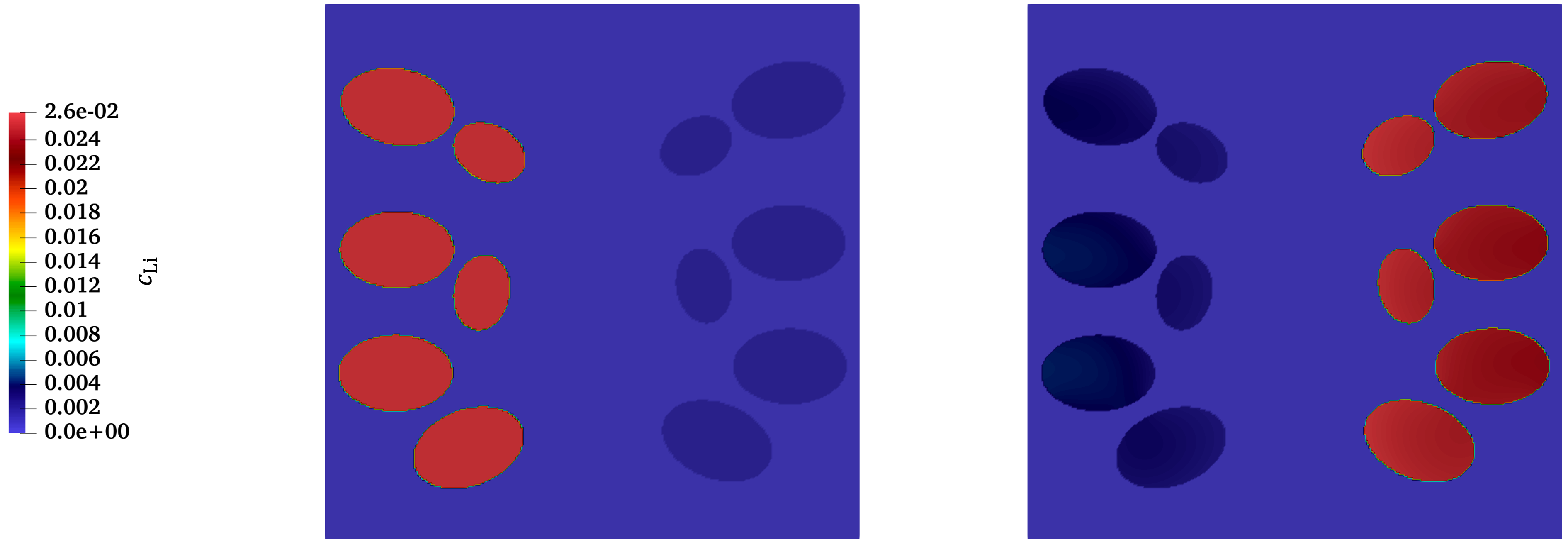}
    \caption{Li concentration field ($\text{pmol/}\mu\text{m}^3$) in a multi-particle cell with stress-dependent kinetics and fracture enabled. Left: initial state. Right: end of first discharge.}
    \label{fig:multiparticleconfig}
\end{figure}
\begin{figure}[h!]
    \centering
    \includegraphics[width=1\linewidth]{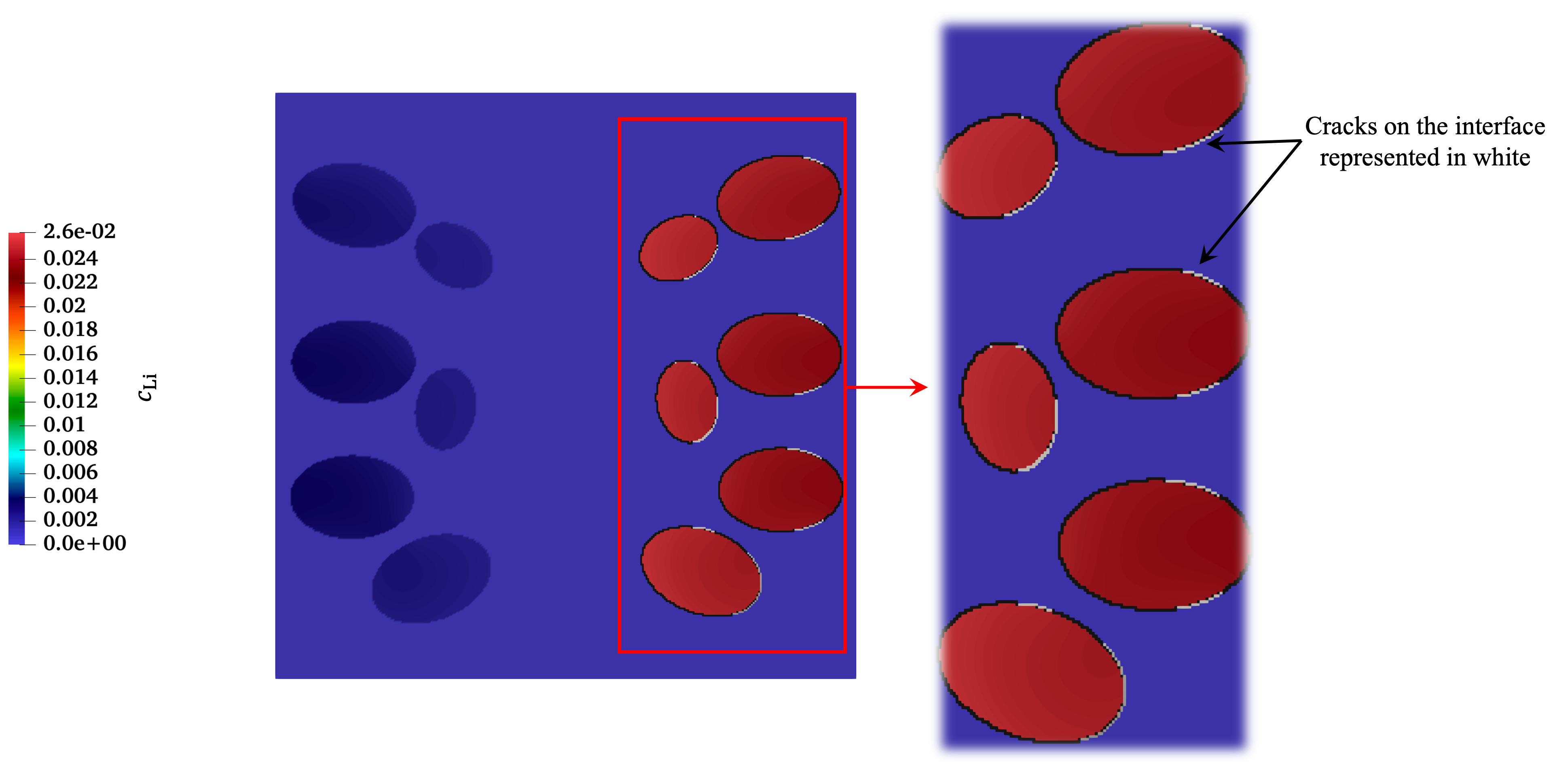}
    \caption{Fractured elements in white on the interface between solid electrolyte and cathode active particles at the end of the first discharge.}
    \label{fig:multiparticlefracture}
\end{figure}
\begin{figure}[h!]
    \centering
    \includegraphics[width=1\linewidth]{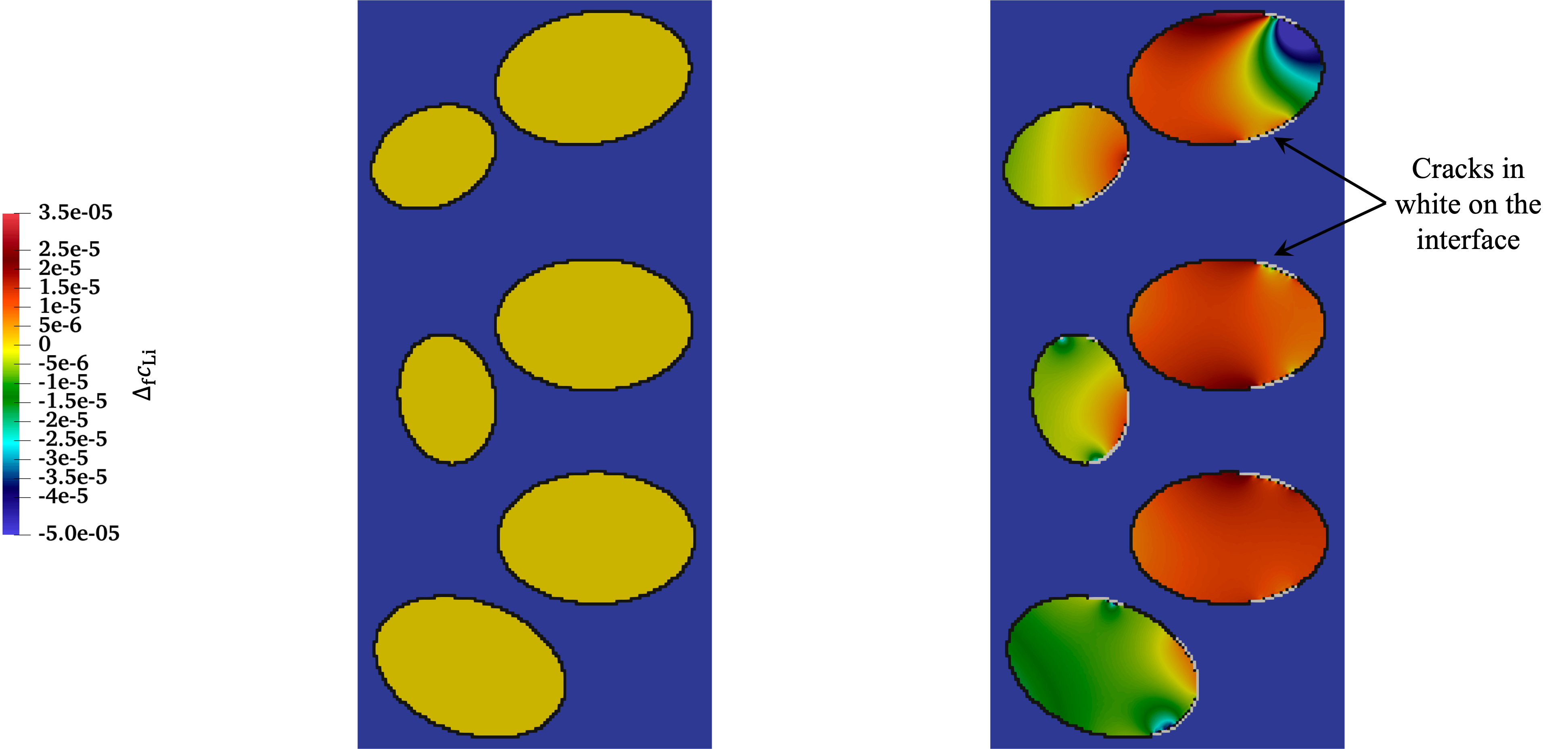}
    \caption{Comparison of the deviation in Li concentration  $\Delta_\text{f} c_{\text{Li}}$ ($\text{pmol/}\mu\text{m}^3$) in cathode active particle between cases with fracture and with fracture suppressed. Left: the end of the first discharge with fracture suppressed, for which $\Delta_\text{f} c_\text{Li} = 0$ by definition. Right: Fracture with cracked elements in white.}
    \label{fig:multiparticlefracturecompare}
\end{figure}

 We demonstrate the robustness of the framework in its extension to modeling the stress-mediated electro-chemo-mechanics including fracture of multiple particles. Fig \ref{fig:material_id} is a multi-particle configuration generated by the workflow in Fig \ref{fig:imagebasedmeshsetup}. Fig \ref{fig:multiparticleconfig} is a detail from a computation on this configuration showing the Li concentration profiles  at the initial state and at the end of the first discharge. Fig \ref{fig:multiparticlefracture} shows the  elements that fractured in white  on the interfaces of cathode active particles during the discharge half-cycle. In addition to the mechanics of each particle, the deformation of the surrounding particles in the multiple-particle cell also enhances the tensile stresses at each particle interface resulting in further fracture  developing in comparison to the single particle cell. To show the degradation in charge transfer across the interfaces of particles as a result of Eq \eref{eq:BV1}, in Figure \ref{fig:multiparticlefracturecompare} we plot the distribution of differences $\Delta_\text{f} c_{\text{Li}}$ similar to the fracture activated simulation in section \ref{sec:singleparticle}. The degradation of Li transfer across the fractured interfaces of each particle is evident.

\section{Discussion}\label{sec:conclusion}

We have presented a contribution to the growing computational treatments of coupled electro-chemo-mechanically driven   degradation of solid state batteries, with a focus on phenomena at the electrolyte-active particle interface.  Key to our treatment is the use of Cartesian meshes with an embedded interface representation, which we regard as a step that will enable an extension of our framework to modelling larger numbers of multi-particle configurations obtained from real battery materials with irregularly shaped particles that also have a distribution of particle sizes. As we have demonstrated, this first necessitated a mathematical treatment of discontinuities in the fields of Lithium and Lithium cation concentrations, and of the electric potential fields at the particle-electrolyte interfaces. It also naturally extends to the treatment of interface fracture via the strong discontinuity treatment. We have focused on interface fracture, motivated by the greater susceptibility to this mode of failure in solid state batteries with stiff ceramic electrolytes and active particles. Also of some interest in this regard is the recent work by Van der Ven et al. on the possible role of ferroelastic toughening mechanisms in preventing interface and intra-particle or electrolyte fracture \cite{van2023ferroelastic}. To enable a coupled solution of the electro-chemo-mechanical equations with interface fracture, we have extended the discontinuous finite element treatment that has appeared previously under the strong discontinuity and variational multiscale frameworks.

We have addressed several aspects of electro-chemo-mechanical coupling: The intercalation strains drive the mechanics with lithiation and delithiation during discharge-charge cycles. The resulting stresses throughout the solid state battery influence the kinetics of diffusion as well as the interface reactions of charge transfer. Additionally, we have accounted for the degradation of charge transfer reactions due to interface fracture and separation of the electrolyte-particle surfaces. Our numerical simulations have demonstrated all these effects: discontinuous concentration and electric potential fields, normal crack opening with fracture, the stress-influenced transport and reaction both in the absence of fracture and with its effect accounted for. 

In this first communication of our framework we have focused on the computational methods and demonstrated the physics that they resolve over a single discharge-charge cycle. The work and results presented here are an early step toward simulating capacity fade over hundreds of cycles driven by the above phenomena and ultimately toward incorporating other coupled electro-chemo-mechanics. 

\section*{Acknowledgements}
We gratefully acknowledge the support of Toyota Research Institute, Award \#849910: ``Computational framework for data-driven, predictive, multi-scale and multi-physics modeling of battery materials''.  This work also used the Extreme Science and Engineering Discovery Environment (XSEDE) Comet at the San Diego Supercomputer Center and Stampede2 at The University of Texas at Austin's Texas Advanced Computing Center through allocation TG-MSS160003 and TG-DMR180072.

\end{document}